\documentclass{aa}  

\usepackage{textcomp}
\usepackage{graphicx}
\usepackage[varg]{txfonts}
\usepackage{mathtools}
\usepackage{flushend}

\usepackage{subcaption}
\usepackage{caption}
\usepackage{siunitx}
\usepackage{hyperref}
\usepackage{lscape}             
\usepackage{placeins} 
\usepackage{xcolor}

\usepackage{graphicx}   
\usepackage{amsmath}    

\usepackage{natbib}
\bibpunct{(}{)}{;}{a}{}{,} 
\begin{document}

    \title{Novel insights into the Coma cluster kinematics with DESI}
    \subtitle{I. Linking mass profile, orbital anisotropy, and galaxy populations}
    \titlerunning{The Coma cluster kinematics with DESI}

    \author{
        S. Pedratti\inst{1}\thanks{E-mail: s.pedratti1@campus.unimib.it},
        L. Pizzuti\inst{1,2},
        M. Fossati\inst{1,3},
        A. Biviano\inst{2,4},
        A. Boselli\inst{5},
        A. Ragagnin\inst{2},
        A. Carlin\inst{1}
        }

    \institute{Dipartimento di Fisica G. Occhialini, Universit\`a di Milano-Bicocca, Piazza della Scienza 3, 20126 Milano, Italy
        \and    
        INAF -Osservatorio Astronomico di Trieste, via G. Tiepolo 11, I-34143 Trieste, Italy
    \and
        INAF - Osservatorio Astronomico di Brera, via Brera 28, 20121 Milano, Italy
    \and IFPU $-$ Institute for Fundamental Physics of the Universe, via Beirut 2, 34014 Trieste, Italy
    \and 
    Aix Marseille Univ, CNRS, CNES, LAM, Marseille, France\thanks{Scientific associate INAF - Osservatorio Astronomico di Cagliari, Via della Scienza 5, 09047 Selargius (CA), Italy}
}

\date{Received 21 January 2026 / Accepted 29 April 2026}

\abstract{We investigate the kinematic properties of the Coma galaxy cluster using a new, large, spectroscopic sample of member galaxies, from the Dark Energy Spectroscopic Instrument (DESI). By means of the \textsc{MG-MAMPOSSt} code, which is based on the Jeans equation, we jointly reconstruct the total cluster mass profile and the velocity anisotropy profile. Assuming a Navarro-Frenk-White model, we estimate a virial mass of $M_{200\rm{c}} = 1.04_{-0.08}^{+0.07}~({\rm stat}) \pm 0.09~({\rm syst}) \times 10^{15} \, \mathrm{M}_\odot $, corresponding to $r_{200\rm{c}}=2.07 \pm 0.05 \, \mathrm{Mpc}$ and a scale radius for the mass profile of $r_{\rm s} =0.73_{-0.30}^{+0.24}\,({\rm stat}) \pm 0.21 \,({\rm syst)} \, \mathrm{Mpc}$, which provides the tightest robust kinematic mass profile constraint to date. By separately considering the mass of the hot gas and the galaxy stellar mass, we further determined the dark matter mass profile, with $M_{200\rm{c}}^{\rm DM} = 8.6^{+1.2}_{-0.8}\times 10^{14}\,\text{M}_\odot$. We discuss the impact of the mass and galaxy number density parametrisations, the effect induced by different choices of the cluster's rest frame and of the radial range of the kinematic analysis, further comparing our results with previous estimates from the literature. The cluster dynamical state was also assessed, using the spatial and line-of-sight velocity distributions of the members.
We also analysed the line-of-sight velocity distributions and anisotropy profiles of different galaxy populations, selected based on their colour (red sequence, green valley, and blue cloud). The orbits of green valley and blue cloud galaxies appear to be more radial in the centre and in the outskirts, respectively, with the latter predicting a higher cluster virial mass. 

This study provides new insights into the interplay between dynamical and intrinsic properties of galaxies in massive structures, which is fundamental to verify the tight connection between a galaxy's evolution and its environment.}


\maketitle
\nolinenumbers

\section{Introduction}
\label{sec:intro}

Galaxy clusters, the largest gravitationally bound systems in the Universe, are excellent laboratories at the crossroads of cosmology and astrophysics. 
In order to employ clusters as reliable cosmological probes, it is essential to use robust methods to estimate their total mass, as well as the radial mass distribution. These can provide fundamental insights into the dark matter properties (e.g. \citealt{Biviano_2023}; \citealt*{Vall_s_P_rez_2025}) and the nature of gravity (e.g. \citealt{Pizzuti22}; \citealt*{Zamani24}). 

In parallel, the deep potential wells of galaxy clusters make them particularly convenient for studying the processes affecting galaxy evolution and feedback scenarios in dense environments (e.g. \citealt{Boselli_2014}; \citealt*{Boselli_2022}; \citealt{Sampaio_2024}). In particular, galaxy populations in clusters are predominantly composed of massive, red, quiescent galaxies with low star formation rates, generally referred to as early-type galaxies (ETGs), alongside a smaller population of low-mass, blue, highly star-forming galaxies known as late-type galaxies (LTGs; \citealt{Dressler_1997, Boselli_2006}).  

The features of galaxies in clusters are determined by a combination of their initial formation conditions and the interactions within the environment in which they evolve. How these two aspects impact the observed population of galaxies in clusters is hotly debated in the astrophysics community. Ram-pressure stripping of the interstellar medium (ISM) is considered to be the dominant mechanism affecting the properties of galaxies in clusters \citep{Boselli_2022, Xie_2025}: as they travel at high velocities (up to $\sim 1000$ km/s) through the hot and dense gas of the intra-cluster medium, the ISM can be stripped away if ram pressure overcomes the gravitational pressure anchoring the gas to the galaxy \citep{Gunn_1972}.
This process is more effective for galaxies on radial orbits, which both increase their speed and are brought closer to the cluster core (\citealt{Vollmer_2001}; \citealt*{Singh_2019}).

Moreover, \citet{Tammann72}, \citet{MD77}, \citet{Sodre+89}, \citet{Biviano+92}, \citet{Colless_1996}, \citet*{ABM98}, and later \citet{Boselli_2006} showed that the line-of-sight (LoS) velocity distributions of galaxies in clusters is type-dependent. While ETGs display a virialised Gaussian velocity distribution, LTGs exhibit strong deviations from Gaussianity, the distribution being generally broader and, in some cases, asymmetric \citep{ZF93,Biviano+97}. These differences in velocity distributions are probably due to LTGs' orbits being more elongated and to their anisotropic infall from the surrounding large-scale filaments,
and serve as an indirect evidence of their infall into the cluster core (e.g. \citealt{Mamon_2019,Pizzuti_2025b}). The more elongated orbits of spiral, gas-rich, star-forming galaxies in dense environments suggest that these galaxies are subject to dynamic gas depletion, mainly due to ram pressure stripping, leading to star formation quenching and morphological transformations of LTGs into ETGs, eventually producing the predominant population of red quiescent systems of present-day clusters.

This paper focuses on the Coma cluster (Abell 1656), one of the most massive and well-studied galaxy clusters of the local Universe at a redshift of $z = 0.023$ \citep{Kang_2025}, whose study is facilitated by its position in the sky near the Milky Way's north pole, where contamination due to dust and gas from our galaxy's plane is negligible. Despite early indications of the presence of subclustering \citep[see][and references therein]{Biviano98}, the Coma cluster has been considered for a long time as the prototype of an old and relaxed cluster. X-ray maps (\citealt*{White_1993}; \citealt{ Neumann_2001}), analyses of the projected distribution of galaxies \citep*{Baier_1990}, and the LoS velocity distribution \citep{Colless_1996} established the existence of a group centred around the massive galaxy NGC 4839, which is now understood to be on secondary infall after reaching its first orbital apocentre \citep[e.g.][]{Biviano+96,Lyskova+19,Churazov_2021}. In addition, a recent or ongoing merger has been identified between the main cluster and a group, partially disrupted, originally dominated by NGC 4889 or NGC 4874 \citep{Biviano+96,Adami_2005a}. Much more recently, \citet{Costa_2025} and \citet{Kang_2025} found evidence of many substructures, often overlooked because of the faint nature of the galaxies tracing them. The interactions that result from this complex configuration could be at the origin of the observed post-starburst (PSB) galaxies on the periphery of the cluster \citep{Caldwell_1993, Gavazzi_2010}.

After \citet{Zwicky33}'s first estimate of the total mass, many studies focused on getting more precise determinations of the total mass distribution through several independent measurements. More specifically, \citet{Lokas_2003} performed a kinematic analysis of elliptical galaxies in the cluster field to obtain a total mass of $M_{200\rm{c}} = 1.2 \times 10^{15} \, M_\odot$\footnote{All quantities presented in this paper are referred to a cosmological model with $H_0=70$ km~s$^{-1}$~Mpc$^{-1}$, $\Omega_m=0.3$, and $\Omega_{\Lambda}=0.7$.} with $30\%$ accuracy, within a corresponding radius of $r_{200\rm{c}}=2.2 \, \mathrm{Mpc}$.\footnote{\citet{Lokas_2003} report values of $M_{100\rm{c}}$ and $r_{100\rm{c}}$, which we converted here to $M_{200\rm{c}}$ and $r_{200\rm{c}}$ following the method described in Appendix~A of \citet*{Tricottet_2025}.} Here and throughout the paper, we refer to  $M_\Delta = \Delta \,H^2(z)/(2G)\times r_{\Delta}^3$ as the cumulative mass enclosed in a radius, $r_{\Delta}$, of a spherical overdensity with an average density $\Delta$ times the critical density of the Universe at the cluster redshift, $\rho_{\rm{c}}(z)$. Following a similar prescription in the literature, we defined quantities at $\Delta = 200$ as the ‘virial’ mass and radius, respectively. 
Weak lensing calculations allowed \citet{Gavazzi_2009} to estimate $M_{200\rm{c}}=5.1^{+4.3}_{-2.1} \times 10^{14} \, M_\odot$, while a recent paper by \citet{Cha_2025} employed a deep learning approach based on a convolutional neural network to mitigate the potential setbacks of traditional lensing analyses, such as noise amplification and mass-sheet degeneracy, finding the value of $M_{200\rm{c}}=7.4^{+2.4}_{-3.1} \times 10^{14} \, M_\odot$. \citet{Mirakhor_2020} combined the electron density profile, recovered from the X-ray observations of \textit{XMM-Newton}, and the pressure profile from the \textit{Planck} Sunyaev--Zeldovich effect~\citep[SZ;][]{Sunyaev1972} observations to estimate the virial mass as $M_{200\rm{c}}=(8.5 \pm 0.55) \times 10^{14} \, M_\odot$, corresponding to a radius of $r_{200\rm{c}}=2.0 \, \mathrm{Mpc}$. Finally, \citet{Ho_2022} made use of recently developed methodologies based on Bayesian deep learning, obtaining $M_{200\rm{c}}=1.8^{+0.74}_{-0.53} \times 10^{15} \, M_\odot$, and, even more recently, \citet{Benisty25} employed the caustics, the virial theorem, and the Hubble-flow methods to find virial masses in the range of $[1.1-2.9]\, \times 10^{15} \,  M_\odot$.

In this work we perform a kinematic analysis of the Coma cluster's member galaxies, using new spectroscopic data collected from the Dark Energy Spectroscopic Instrument \citep[DESI;][]{DESI_DR1} to jointly reconstruct the mass and anisotropy profiles under the assumption that the cluster is in an equilibrium configuration. We employ the \textsc{MG-MAMPOSSt} code of \citet*{Pizzuti_2021}, which applies the Jeans equation to the projected positions and LoS velocity distribution of member galaxies. 
We study the relation between the Coma cluster member galaxies and the global cluster environment, by tracing the behavior of the orbits for different subsamples of galaxies based on their colour (red sequence, green valley, and blue cloud), further discussing the impact of systematics due to mass and anisotropy models and different  sample parameters (e.g. projected radial ranges, choice of the centre, velocity rest frame).
Our findings provide the most stringent constraints on the mass distribution in Coma and on the orbits of its member galaxies to date.

The work is structured as follows. Section~\ref{methods} describes the DESI dataset of galaxies in the Coma cluster and the preliminary analysis performed on the spatial and velocity distribution of galaxies. In Sect.~\ref{sec:mamposst} we provide an overview of the \textsc{MG-MAMPOSSt} method, employed to study the kinematics of the system, and of the various models adopted in the analysis. In Sect.~\ref{results} we show our results, further discussing the impact of systematic effects and checking the robustness of our analysis. The conclusions and final considerations are reported in Sect.~\ref{sec:conclusions}.

\section{Dataset and preliminary analyses}
\label{methods}

\subsection{Dataset}
\label{subsec:dataset}

Our dataset comes from the first Data Release (DR1) of DESI \citep{DESI_DR1}, a highly multiplexed spectroscopic instrument mounted on the Mayall 4-metre telescope at the Kitt Peak National Observatory. The Coma cluster was targeted by Science Verification (SV) programmes aimed at obtaining a high density of cluster targets. As a result, the final target density in the area is much larger than the average of the DESI survey, providing us with more than 2,000 spectra in a DESI pointing. We removed all the spectra from non-nuclear sources or misclassified stars, through visual and kinematic inspection, respectively, and thus obtained a pure sample of nuclear spectra of galaxies.
We required the objects to have at least $g$- and $r$-band magnitudes ($m_g$ and $m_r$, respectively) for colour selections. Information on access to the resulting data is provided in the \hyperref[sec:data_availability]{data availability} section. We further selected only galaxies with $m_r <19.7$ mag to ensure a high spectroscopic completeness of the photometrically selected sample. We estimated the spectroscopic completeness by selecting all the extended (galaxy-like) photometric objects brighter than the magnitude cut, deriving the fraction that has available DESI spectroscopy in bins of distance from the centre of the Coma cluster, and excluding the central 100 kpc, as is further discussed below. We find an overall spectroscopic completeness of 91\% in our sample, with no significant radial trends (see Appendix \ref{app:completeness}).

\begin{figure}
    \centering
    \includegraphics[width=0.85\linewidth]{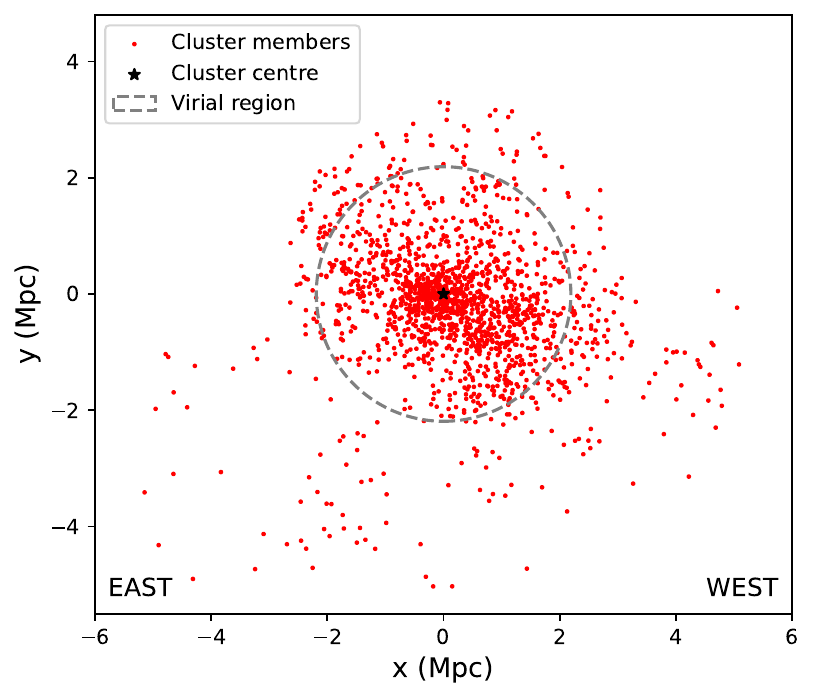}
    \caption{Projected positions of galaxies in the Coma cluster. The point marked with a black star indicates the cluster centre, whose co-ordinates (RA=194.90 deg, DEC=27.96 deg) have been taken from \citet{Malavasi20} as the position of NGC 4874, while the grey circle encloses the region within $r_{200\rm{c}} = 2.19 \, \mathrm{Mpc}$, quoted in \citet{Boselli_2006}. }
    \label{fig:full_distrib_DESI}
\end{figure}

\begin{figure}
    \centering
    \includegraphics[width=0.9\linewidth]{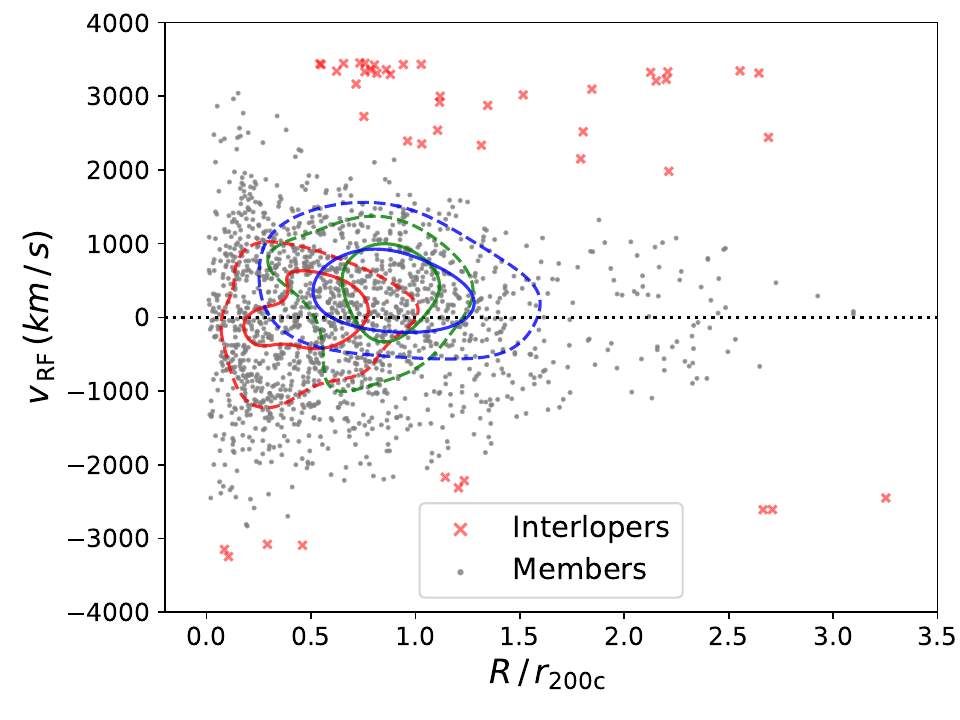}
   \caption{Projected phase-space distribution of member galaxies in the Coma cluster (grey dots) and interlopers removed by Clean (red crosses). The dashed black line indicates the average velocity of the cluster (Hubble + bulk motion), while coloured curves highlight iso-density contours (dashed and solid lines correspond to 90th and 97th percentiles, respectively) for RS (red), GV (green), and BC (blue) galaxies. The projected distance from the centre is normalised to $r_{200\rm{c}}$, taken equal to the value quoted in \citet{Boselli_2006}, $2.19 \, \mathrm{Mpc}$.}
    \label{fig:phase_space_DESI}
\end{figure}

Our final spectroscopic galaxy catalogue includes 1758 galaxies and covers the angular co-ordinate range (J2000) $190^\circ < \text{RA} < 200^\circ$ and $25^\circ < \text{DEC} < 31^\circ$, for a total sky area of $\sim 53 \, \deg^2$ around the Coma cluster region, extending out to $\sim 3\,r_{200\rm{c}}$. We consider the cluster centre to be located at (RA = $194.90 \,\deg$, DEC = $27.96 \,\deg$), corresponding to the  position of the elliptical galaxy NGC 4874 (see e.g. \citealt{Malavasi20}). In order to avoid systematics in the kinematic analyses, it is essential to remove the interlopers, i.e. galaxies that are within the virial radius in projection but outside the virial sphere. This step has been performed using the Clean algorithm \citep*[Appendix B of][]{Mamon_2013}, which provides a more refined approach with respect to a simple phase-space cut. Figure~\ref{fig:full_distrib_DESI} depicts the projected positions of the 1651 galaxies in our final Coma cluster sample. 
As we discuss in Sect.~\ref{sec:mamposst}, our kinematic analysis requires the projected positions, $R$, from the cluster centre and the LoS velocities of the member galaxies, $v_\text{RF}$, computed in the cluster's rest frame.  The space of pairs ($R,v_\text{RF}$) is called projected phase-space (p.p.s.), shown in Fig.~\ref{fig:phase_space_DESI}. 

In order to study the orbital properties of different galaxy populations, we split the sample into three main categories according to their position in the $(m_g-m_r)$ versus stellar mass ($M_*$) diagram. The stellar masses were computed from the DESI photometry using CIGALE \citep{Boquien2019} SPS modelling with exponentially declining SFHs, BC03 models, and a \cite{Chabrier2003} initial mass function.
 
The resulting classification into red sequence (RS), green valley (GV), and blue cloud (BC) galaxies follows the representation chosen by \citet{Boselli_2014} to study the Virgo cluster and takes into account the different ages of galaxies. We started by visually identifying the RS galaxies and fitting them with a straight line; then, we assigned galaxies to the GV 
if they lie between $2\,\sigma$ and $5\,\sigma$ from the RS, where $\sigma$ is the dispersion of 
the RS ($m_g-m_r$) colours at a given mass.
We therefore considered RS galaxies to be those with $m_g-m_r>0.07 \,\log(M_*/M_\odot) + 0.03$, BC galaxies those with $m_g-m_r<0.07\, \log(M_*/M_\odot) - 0.07$, and GV those in between. While the $m_g-m_r$ colour does not offer a large wavelength baseline, and the GV classification is not unique, our results are not affected by the exact choice of the GV location, given the small number of galaxies in this region.

\begin{figure}
    \centering
    \includegraphics[width=0.9\linewidth]{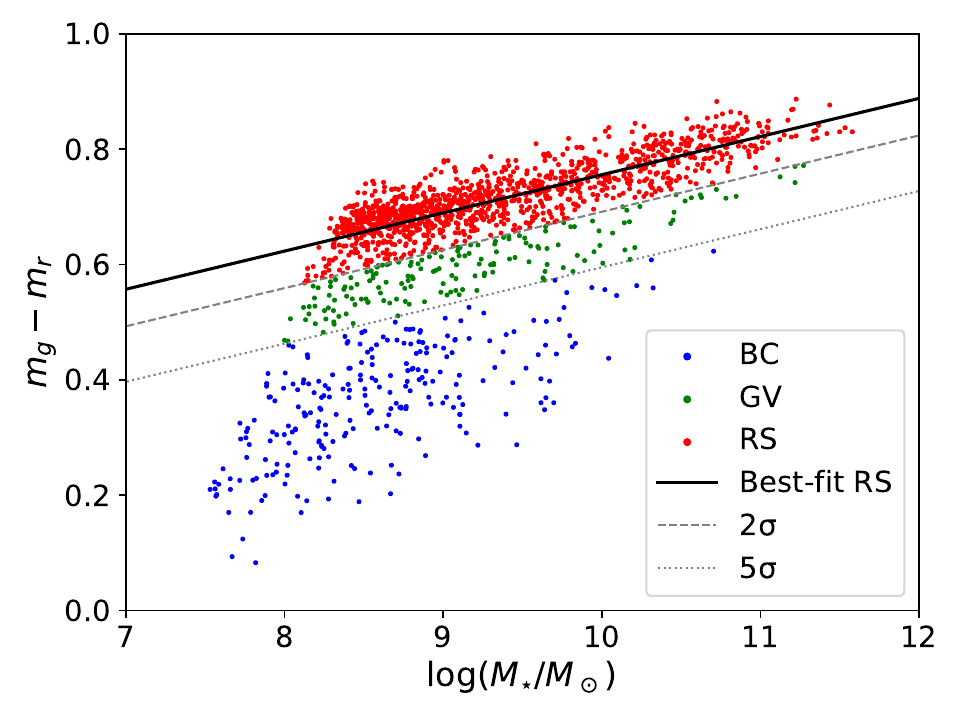}
   \caption{$(m_g-m_r)$ vs stellar mass diagram. The colour of the markers highlights the three colour classes: BC in blue, RS in red, and GV in green. The solid line represents the best fit for RS galaxies. The dashed and dotted lines indicate the $2\,\sigma$ and $5\,\sigma$ limits, respectively.}
    \label{fig:colors}
\end{figure}

Finally, 19 galaxies were removed from the classification due to unreliable photometry. The resulting subsamples consist of 1216 RS, 167 GV, and 249 BC galaxies, for a total of 1632 galaxies, whose $(m_g-m_r)$ versus $M_*$ diagram is shown in Fig.~\ref{fig:colors}.

\subsection{Galaxy number density profile}
\label{num_density_methods}

The distribution of galaxies in clusters is shaped by that of dark matter, which dominates the total mass of the system and determines the overall gravitational potential.  
A widely adopted model to describe the total mass density profiles of gravitationally bound systems, such as galaxy clusters, is the  Navarro-Frenk-White (NFW) model \citep*{Navarro_1997}:

\begin{equation}
    \rho_{\rm NFW}(r)=\frac{\rho_{0,_{\rm NFW}}}{(r/r_{\rm s})\, (1+r/r_{\rm s})^2} \, ,
    \label{NFW}
\end{equation}
proportional to $r^{-1}$ in the innermost region and to $r^{-3}$ at large radii. In Eq.~\ref{NFW}, $\rho_0$ is the characteristic density of the cluster and $r_{\rm s}=r_{-2}$ is the scale radius at which the logarithmic derivative of the profile equals $-2$. While initially proposed as a description of dark matter halos in N-body cosmological simulations, the NFW profile has been shown to provide an adequate description of the total mass distribution of clusters (e.g. \citealt{Umetsu16, Pizzuti_2025b}).

The surface density profile predicted for the NFW model \citep{bartelmann1996} is also suitable for the distribution of the projected radii of galaxies, $\nu(r)$ (e.g. \citealt*{Lin_2004}; \citealt{Annunziatella_2014}; \citealt{Sartoris_2020}); however, it is important to note that the scale radius of the galaxy number density, $r_\nu^{\rm NFW}$, is not in general the same as the scale radius, $r_{\rm s}$, of the mass profile \citep{Biviano_2006, Budzynski_2012}. In fact, while RS galaxies typically follow a similar distribution to that of dark matter, BC galaxies tend to have a much less concentrated profile, with typical values of $r_\nu^{\rm BC} \sim 4 \, r_\nu^{\rm RS}$ (e.g. \citealt{Collister_2005}).

We further considered the Hernquist model (and its projection, \citealt{Hernquist_1990}), which exhibits a faster decline at large radii:
\begin{equation}
    \nu_{\rm H}(r) \propto \frac{1}{(r/r_{\nu}^{\rm H})\, (1+r/r_{\nu}^{\rm H})^3} \, ,
    \label{Hernquist}
\end{equation}
with the characteristic scale of $r_\nu^{\rm H}=2\,r_{-2}$, so that the comparison of $r_{-2}$ between Hernquist and NFW models implies $r_\nu^{\rm H}=2\,r_\nu^{\rm NFW}$.

We performed a maximum likelihood fit of the parameters of the surface density profiles of both NFW and Hernquist models to the distribution of projected radii of galaxies. This avoided the need to split the projected radii in bins \citep{Sarazin_1980, Biviano_2013}, assuming a constant completeness of the sample in the projected radial distance explored. 

\begin{figure}
    \centering
    \includegraphics[width=0.9\linewidth]{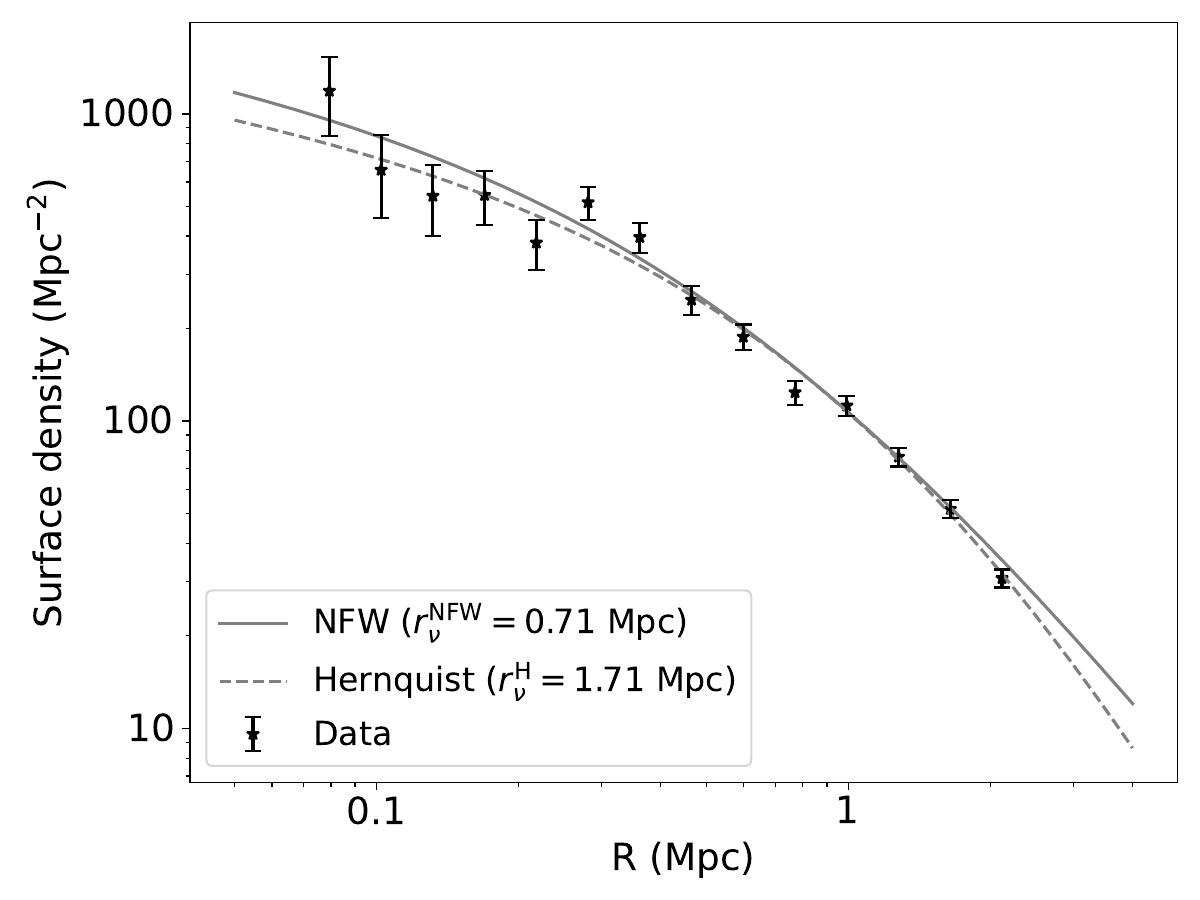}
    \caption{Surface density profile of the Coma cluster fitted with NFW (solid line) and Hernquist (dashed line) models. The error bars correspond to the Poisson uncertainties.}
    \label{fig:num_density_pNFW}
\end{figure}

The best-fit values obtained are $r_\nu^{\rm NFW} = 0.71^{+0.07}_{-0.06} \, \mathrm{Mpc}$ in the case of NFW and $r_\nu^{\rm H} = 1.71^{+0.11}_{-0.06} \, \mathrm{Mpc}$ with Hernquist. Both models match the observed number density distributions adequately well, with no significant differences; as such, during the rest of this work we assume a Hernquist model as the reference model for the galaxy number density profiles.
The validity of this choice is assessed in Appendix~\ref{app:num_density_assump}, where we compare the outcome of \textsc{MG-MAMPOSSt} runs with NFW and Hernquist models and we treat the scale radius, $r_\nu^{\rm NFW}/r_\nu^{\rm H}$, as both a free and a fixed parameter. Figure~\ref{fig:num_density_pNFW} shows the surface density profile of galaxies fitted with both profiles.

Since in this work we analyse different subsamples of galaxies based on their colours, we further performed a fit of the number density profiles of RS, BC, and GV galaxies separately, assuming a Hernquist model (see Fig. \ref{fig:surface_density_subsamples}), finding the scale radii of $r_\nu^{\rm RS}=1.28^{+0.04}_{-0.08} \, \mathrm{Mpc}$,
$r_\nu^{\rm BC}=5.82^{+3.49}_{-1.80} \, \mathrm{Mpc}$, and $r_\nu^{\rm GV}=12.79 \, \mathrm{Mpc}$ with a lower $1\,\sigma$ uncertainty of $5.71 \, \mathrm{Mpc}$, while the upper bound remains unconstrained due to the particularly flat distribution of GV galaxies in our sample.  It is worth noticing that these results are consistent with the expectation of $r_\nu^{\rm BC} \sim 4 \, r_\nu^{\rm RS}$ (e.g. \citealt{Collister_2005}). 

Cutting the LoS velocities in p.p.s. does
not remove interlopers along the LoS that have small absolute LoS velocities \citep*[e.g.][]{Mamon_2010}. We therefore also tried fitting the number density distributions of the three classes by adding a constant field as an additional free parameter. The resulting best-fit $r_{\nu}$ values are within the 1 $\sigma$ error bars of the values obtained without the constant field. The difference can be quantified by using the BIC criterion, defined as
\begin{equation}
    \text{BIC} = k \, \ln(N) - 2 \, \ln\mathcal{L} \, ,
    \label{eq:BIC}
\end{equation}
where $k$ is the number of free parameters in the run, $N$ is the total number of tracers used in the analysis, in our case the member galaxies, and $\ln\mathcal{L}$ is the best-fit log-likelihood. Lower BIC values indicate preferred models: in fact, the penalty term, $k \, \ln(N)$, increases with both the number of parameters and the number of data points, discouraging overly complex models unless they provide a significantly better fit. As $k$ increases, the model gains flexibility and typically achieves a higher (less negative) log-likelihood, but this comes at the cost of a larger penalty. Conversely, a model with fewer free parameters is rewarded by a smaller penalty. We find $\text{BIC}_{\rm no\, field} - \text{BIC}_{\rm field} = -6.9, -4.9, -1.0$, for the RS, GV, and BC samples, respectively, which indicates that adding an additional free parameter is not statistically warranted. In the following we proceed with the $r_{\nu}$ values obtained by the fit with a single free parameter.

\subsection{LoS velocity distributions}
\label{subsec:LoS_methods}

As a first diagnostic of the dynamical state of the system, we studied the deviations from Gaussianity of the LoS rest-frame velocity distribution of member galaxies \citep{Pizzuti_2020}. This indicator strongly correlates with other commonly used relaxation diagnostics based on X-ray observations, reinforcing its suitability for characterising the global dynamical state of a cluster \citep*{Roberts_2018}. As was mentioned in Sect.~\ref{sec:intro}, observations and simulations consistently show that in dynamically relaxed systems the distribution of LoS velocities are nearly Gaussian, whereas unrelaxed or disturbed systems exhibit significant departures from Gaussianity \citep{Yahil_1977, Boselli_2006, Ribeiro_2013}. To quantify such deviations, we applied the Anderson-Darling test \citep{Anderson_1952} to the LoS velocity distributions for RS, GV, and BC subsamples, up to $3$ Mpc. Values of the Anderson-Darling distance $A^2$ coefficient that are above $0.752 \,(1.035)$ indicate a $\geq 95\,(99)\%  $ confidence that the velocity distribution is inconsistent with a Gaussian.

\begin{figure}
    \centering
    \includegraphics[width=0.8\linewidth]{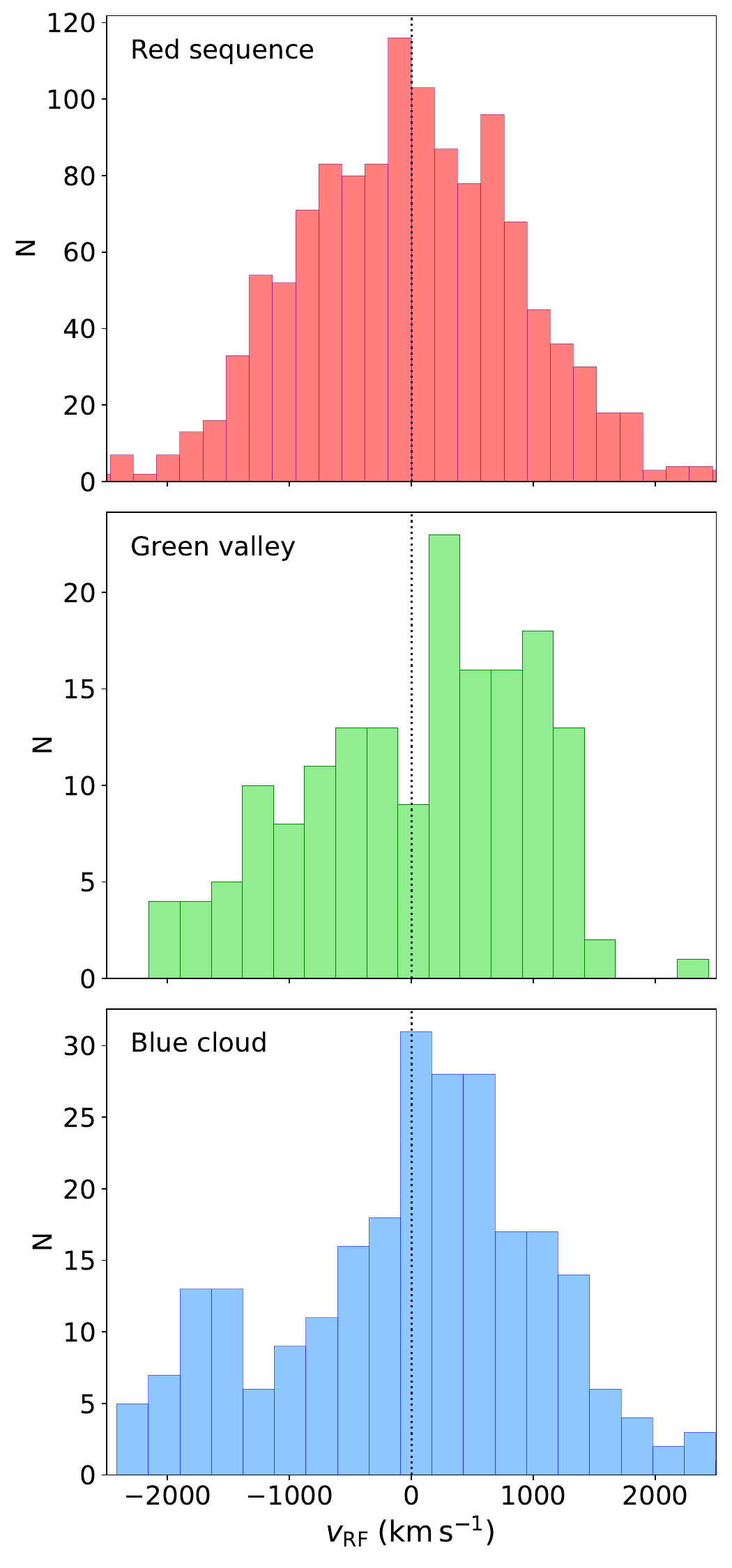}
    \caption{LoS velocity distributions of RS (\textit{top}), GV (\textit{middle}), and BC (\textit{lower}) galaxies. The vertical dashed line indicates the average velocity of the cluster. }
    \label{fig:vel_LoS}
\end{figure}

The LoS velocity distributions for RS, GV, and BC galaxies are shown in Fig.~\ref{fig:vel_LoS}. The three distributions are not significantly different, according to a Kruskal-Wallis test \citep{KW52}, nor are their medians, according to two-by-two comparisons by the Sign test \citep{DM46}. However, the three distributions display different shapes.
The resulting $A^2$ coefficient for RS galaxies is $0.29$, a low value that suggests that orbits are almost completely virialised. On the contrary, BC galaxies exhibit a notable deviation from Gaussianity with $A^2 = 1.56$, corresponding to a rejection confidence of $99\%$, due to a skewed distribution towards higher velocities. The excess of high velocities for BC galaxies is clearly visible in Fig.~\ref{fig:phase_space_DESI} at $R<r_{200\rm{c}}$, and it causes a
negative correlation between $R$ and $v_\text{RF}$ (Spearman rank correlation coefficient $r=0.097$, corresponding to a probability of 0.05). No such correlation is present for the RS and GV classes. These characteristics of the BC velocity distribution suggest more elongated orbits for LTGs, which might be infalling towards the cluster centre from a filamentary-like structure at the near-side of the cluster along the LoS. This interesting feature will be addressed specifically in an upcoming paper (Paper II), dedicated to substructures and dynamical state, along with other possible criteria based on higher moments of the velocity distribution that may help to refine the assessment of the cluster equilibrium (or lack of it). 
The GV galaxies' velocity distribution shows a quite unrelaxed behaviour too, with a coefficient of $A^2 = 1.27$, rejecting Gaussianity again at $99\%$ confidence. As for the full sample, we found a value of $A^2 = 0.80$, higher than that of the RS but smaller than that of the GV, which indicates that the overall system is quite far from a full dynamical relaxation state (rejected Gaussianity at $95\%$ confidence), consistent with the presence of still infalling star-forming galaxies. Similar LoS velocity distributions of Coma galaxies, distinguished by morphological type, were previously reported by \citet{Andreon_1996}, based on a much smaller sample.

\section{Kinematic analysis with MG-MAMPOSSt}
\label{sec:mamposst}

We performed a kinematic analysis of the Coma cluster member galaxies by means of the \textsc{MG-MAMPOSSt} (Modified Gravity - Modelling Anisotropy and Mass Profiles of Observed Spherical Systems) code of \citet{Pizzuti_2021}, an extended version of the \textsc{MAMPOSSt} algorithm of \cite{Mamon_2013}, which handles a wide variety of cosmological set-ups (including $\Lambda$CDM and several modified gravity and dark energy frameworks). \textsc{MAMPOSSt} is a way to jointly reconstruct the anisotropy and mass profiles of spherical systems using the kinematics of the system's members; in particular, the code assumes dynamical relaxation, spherical symmetry, and a shape for the three-dimensional velocity distribution of the tracers of the gravitational potential (i.e. cluster member galaxies). Taking as an input the p.p.s. dataset $(R,v_\text{RF})$,  \textsc{MAMPOSSt} performs a maximum likelihood fit to solve the stationary and spherically symmetric Jeans equation:

\begin{equation}
    \frac{\text{d}[\nu(r) \, \sigma_r^2(r)]}{\text{d}r} + 2\beta(r) \frac{\nu(r) \, \sigma_r^2(r)}{r} = - \nu(r) \frac{\text{d}\Phi(r)}{\text{d}r} \, ,
    \label{eq:jeans}
\end{equation}

\noindent where $\nu(r)$ is the galaxy number density profile, $\sigma_r(r)$ is the velocity dispersion in the radial direction, $\beta(r)$ is the orbital anisotropy profile, and $\Phi(r)$ is the gravitational potential, linked to the dynamical mass profile through the Poisson equation:
\begin{equation}
    \frac{\text{d}\Phi}{\text{d}r} = \frac{G \, M_{\text{dyn}}(r)}{r^2} \, .
\end{equation}
The orbital anisotropy profile, $\beta(r)$, is defined as

\begin{equation}
    \beta(r) = 1 - \frac{\sigma_\theta^2 + \sigma_\phi^2}{2 \,\sigma_r^2} \, ,
\end{equation}
where  $\sigma^2_\theta$ and $\sigma^2_\phi$ are the velocity dispersions in the tangential and azimuthal directions, respectively. This reduces to 

\begin{equation}
    \beta(r) = 1 - \frac{\sigma_\theta^2}{\sigma_r^2} 
    \label{eq:beta_profile}
\end{equation}
in the case in which the cluster velocity structure is invariant under rotation about its center, for which $\sigma^2_\theta =\sigma^2_\phi$. Negative values of $\beta(r)$ indicate tangentially biased orbits, while positive values signify a preference for radial orbits. A value close to $\beta = 0$ indicates isotropic orbits, typical of central regions of dynamically relaxed systems. Previous studies based on observations \citep[e.g.][]{Host_2009, Biviano_2013} and simulations \citep[e.g.][]{Hansen_2006, Mamon_2010} of galaxy clusters indicate a growing trend with radius for anisotropy, from central isotropy. 

\textsc{MAMPOSSt} estimates the probability, $q\, (R_i, v_i^\text{RF} \, | \, \vec{\theta})$, of
observing the \textit{i}-th tracer of the gravitational potential (e.g. a cluster galaxy) at a projected distance from the cluster centre ($R_i$) and LoS rest-frame velocity ($ v_i^\text{RF}$), given the set of parameters, $\vec{\theta}$, defining the models for $M_{\text{dyn}}(r)$ and $\beta(r)$. According to \citet{Mamon_2013}:
\begin{equation}
    q\, (R, v_{RF}) = \frac{2\pi R \, g\, (R, v_{RF})}{N_\text{proj}(R_\text{max})-N_\text{proj}(R_\text{min})} \, ,
\end{equation}
where $N_\text{proj}(\,R\,)$ is the predicted cumulative number of galaxies within projected radius, $R$, from the centre, obtained by integrating within the circle of radius $R$ the surface density, $\Sigma(R)$ (i.e. the integral of $\nu(r)$ along the LoS). $R_\text{min}$ and $R_\text{max}$ are, respectively, the minimum and maximum projected radii considered in the p.p.s. and

\begin{equation} \label{eq:probdens}
    g(R,v_{RF}) = \sqrt{\frac{2}{\pi}} \int_R^\infty \frac{r \,\nu(r)}{\sqrt{r^2 - R^2}} \frac{1}{\sigma_z(r, R)} \exp\left[ -\frac{(v_{RF})^2}{2 \sigma_z^2(r, R)} \right] \, \text{d}r \, ,
\end{equation}
where $\sigma^2_z(r,R) = \left[ 1 - (R^2/r^2)\,\beta(r) \right]\sigma^2_r(r)$ is the velocity dispersion along the LoS.

Other similar approaches have been proposed that use additional information from the observed kinematics beyond the simple velocity dispersion profile; for example, the kurtosis profile \citep{Lokas02}. The output of the code is the best-fit estimate of the parameters $\vec{\theta}$ and the corresponding value of $-\ln \mathcal{L}$, computed as

\begin{equation}
    -\ln \mathcal{L} = -\sum_{i=1}^{N} \ln  q\, (R_i, v_
{i}^\text{RF} \, | \, \vec{\theta})  \, ,
\end{equation}
where $N$ is the total number of tracers.
The current implementation of \textsc{MG-MAMPOSSt} assumes a 3D Gaussian velocity distribution; despite its simplicity,  this choice has been proven to also be valid when the LoS velocity distribution significantly deviates from Gaussianity \citep{Mamon_2013, Read21}. Refer to \citet{Mamon_2013} for all the expressions regarding the \textsc{MAMPOSSt} procedure.

For our kinematic analysis,  we considered a total sample (i.e. BC + RS + GV) of 1324 galaxies out of 1651 members, within a projected radius of $R \ge 0.10 \, \mathrm{Mpc}$ and $R \le 2.30 \,\mathrm{Mpc} $ (chosen as $1.1$ times our initial guess for $r_{200\rm{c}}$) from the cluster centre. We excluded the innermost region, as done in many similar works, where cluster dynamics is dominated by the BCG and the mass profile, $M_{\rm dyn}(r)$, is likely to deviate from simple profiles that are known from cosmological simulations to adequately describe the distribution of dark matter (e.g. NFW).
Moreover, in Coma the core is dominated by two bright galaxies that are 200 kpc apart on the sky, namely NGC 4874 and NGC 4889, which  would need a non-trivial modelling of the mass profile.

As for the external regions, at large radii, galaxies are likely to belong to surrounding large-scale structures \citep[e.g.][]{Malavasi20}, and are not gravitationally bound to the cluster, so that they cannot be used as tracers of the cluster potential in the dynamical equilibrium configuration. To ensure the robustness of the mass and anisotropy reconstruction, we performed additional runs of \textsc{MG-MAMPOSSt} varying the inner and outer radius, changing the centre and the velocity rest frame of the cluster. The results of these tests are reported in Sect.~\ref{subsec:syst}.

We explored four different mass models for the total mass of the Coma cluster in the kinematic analysis. We employed the classical NFW (see Eq. \ref{NFW}), whose total mass can be written as 
\begin{equation}
         M_\text{NFW}(r) = M_{200\rm{c}}\frac{\text{ln}(1+r/r_\text{s}) - \frac{r/r_\text{s}}{(1+r/r_\text{s} )}  }{\text{ln}(1+c_{200}) - \frac{c_{200}}{(1+c_{200} )}  }\,,
        \end{equation}
where $c_{200} = r_{200\rm{c}}/r_{\rm s}$ is called ‘concentration’, together with its generalised version, gNFW~\citep*{Nagai2007}:

\begin{equation}
    \small
    \begin{aligned}
    M_{\text{gNFW}}(r)=\; & M_{200\rm{c}} \, (r/r_{200\rm{c}})^{3-\alpha} \\ & \times  \frac{\prescript{}{2}F_1(3-\alpha,3-\alpha,4-\alpha,-r/r_{\rm s})}{\prescript{}{2}F_1(3-\alpha,3-\alpha,4-\alpha,-r_{200\rm{c}}/r_{\rm s})} \, ,
    \end{aligned}
    \label{gNFW}
\end{equation}
where $\prescript{}{2}F_1(a,b,c,x)$ is the ordinary (Gaussian) hypergeometric function \citep[e.g.][]{Mamon_2019}. This model adds an extra free parameter, $\alpha$, that controls the slope. Note that, instead of the characteristic density, $\rho_0$, we considered the virial radius, $r_{200\rm{c}}$, as a free parameter of all the profiles. 

The Hernquist \citep{Hernquist_1990} model,
    \begin{equation}
         M_\text{H}(r) = M_{200\rm{c}}\frac{(r_{200\rm{c}}+ r_\text{H})^2}{r_\text{200}^2}\frac{ r^2}{(r+ r_\text{H})^2}\,,\label{eq:her}
    \end{equation}
and the Burkert \citep{Burkert_1995} model,
    \begin{equation}
    \small
         M_\text{\rm B}(r) = M_{200\rm{c}}\frac{\text{ln}\big[1+(r/r_\text{B})^2\big] +  2\,\text{ln}(1+r/r_\text{B}) - 2\,\text{arctan}(r/r_\text{B}) }{\text{ln}\big[1+(c_{\rm B})^2\big] +  2\,\text{ln}(1+c_{\rm B}) - 2\,\text{arctan}(c_{\rm B})}\,,\label{eq:bur}
    \end{equation}
have been further adopted in the analysis. In Eq.~\eqref{eq:her}, $r_{\rm H} = 2 \, r_{-2}$, while in Eq.~\eqref{eq:bur}, $r_{\rm B} \simeq 0.657 \, r_{-2}$ and $c_{\rm B} = r_{200\rm{c}}/r_{\rm B}$. The Burkert profile has the same asymptotic behaviour as NFW but exhibits a core in the centre ($\rho \propto \mathrm{const}$) rather than a cusp, and a sharper transition from slope 0 to slope 3 than NFW. We denote with $r_\rho$ the general scale radius, which takes the form of $r_{\rm s}$, $r_{\rm H}$, or $r_{\rm B}$ depending on the mass model used.

For the anisotropy profile, we considered two general models; namely, a generalised version of the  
\citet[][gT hereafter]{Tiret+07} profile,

\begin{equation}
    \beta_{\rm gT}(r)=\beta_0 +(\beta_\infty-\beta_0)\, \frac{r}{r +r_\beta} \, ,
    \label{Tiret}
\end{equation}

\noindent and the generalised Osipkov-Merritt (gOM) model \citep{Osipkov_1979, Merritt_1985},

\begin{equation}
    \beta_{\rm gOM}(r)=\beta_0 +(\beta_\infty-\beta_0)\, \frac{r^2}{r^2 +r_\beta^2} \, .
    \label{Osipkov-Merritt}
\end{equation}
where $\beta_0$ and $\beta_\infty$ are the anisotropies at the centre and large radii, respectively, and $r_\beta$ is a characteristic radius. These profiles capture a large variety of orbits from the innermost region to the outskirts of a cluster. Note that, as has been done in previous works (e.g. \citealt{Biviano24}), we assumed $r_\beta= r_{-2}$ of the mass profile, a reasonable assumption to avoid an extra free parameter in the fit, which, in our case, would be largely unconstrained.

The high-quality performance of the DESI instrument results in very small uncertainties on the LoS velocities (see \citealt{DESI_DR1}), which are $<10$ km/s for emission line galaxies and between 5-30 km/s for passive galaxies, as a function of magnitude. In our analysis we adopted a constant uncertainty of $40 \, \mathrm{km \, s^{-1}}$, as a conservative safe upper limit for all the galaxies in our sample.

\textsc{MG-MAMPOSSt} is equipped with a Monte Carlo-Markov chain (MCMC) module to efficiently explore the parameter space, based on a simple Metropolis-Hastings algorithm with a fixed-step Gaussian random walk; the code can handle both flat and Gaussian priors on the free parameters. In the following, we sample 110\,000 points and discard the first 10\,000 as the burn-in phase. 
This number of steps is sufficient to ensure the convergence of the chains; checks have been performed by previous studies in similar frameworks \citep{Pizzuti_2024,Biviano25}, in which the Gelman-Rubin diagnostic coefficients, $\hat{R}$ \citep{Gelman_1992}, have been computed after performing  chains on each run, satisfying the requirement of $\hat{R}<1.1$. Each chain takes $\lesssim 1 $ hour to be completed.
The details of the analysis and the results of \textsc{MG-MAMPOSSt} for the various subsamples of member galaxies are presented and discussed in the next section.

\section{Results}
\label{results}

\subsection{Mass and anisotropy profile analysis}
\label{full_sample_mamposst}

The strategy we adopted to robustly model the mass and orbital anisotropy profiles of the Coma cluster is the following: first, we performed a run of \textsc{MG-MAMPOSSt} on the p.p.s. of the mixed sample of 1324 galaxies, using all the RS, GV, and BC populations as tracers of the gravitational potential. This methodology has been used several times in the literature and it has been shown to work quite well when comparing to independent mass estimates \citep[e.g.][]{Sartoris_2020, Biviano_2013, Biviano_2023}. Then, we split the sample by colour class and we considered each population as an independent tracer of the total gravitational potential \citep[e.g.][]{Mamon_2019}, running \textsc{MG-MAMPOSSt} to sample the population-joint likelihood (Eq. \ref{eq:joint}). 
This procedure allowed us to check whether the two independent Bayesian estimates are in agreement with each other, and provide further support to the viability of mixing different galaxy populations to reconstruct the total gravitational potential of clusters.

We considered two free parameters for the mass profile, $r_{200\rm{c}}$ and $r_\rho$.
We also treated the scale radius of the galaxy number density profile, $r_\nu^{\rm H}$, as a free parameter, assuming a Hernquist model (see Appendix \ref{app:num_density_assump}). In addition, letting $\mathcal{A}=\sigma_r /\sigma_\theta = (1-\beta)^{-1/2}$, we adopted two free parameters for the velocity anisotropy: $\mathcal{A}_0$ and $\mathcal{A_{\infty}}$ for the inner and outer values, respectively. This definition rescales the range $-\infty < \beta < 1$ into $0 < \mathcal{A} < \infty$, preventing numerical issues at $\beta \sim 1$.  
The assumed priors are listed in Table \ref{tab:priors_flat}: they are flat for every parameter and uninformative for each of them except for $r_\nu^{\rm H}$, for which we adopted informative priors based on the best-fit values discussed in Sect. \ref{num_density_methods}.

\begin{table}
\centering
\renewcommand{\arraystretch}{1.1}
\caption{Assumed priors for \textsc{MG-MAMPOSSt} runs with mixed components.}
\begin{tabular}{lccccc}
\hline
{Parameter} & {Prior type} & {Min} & {Max} \\
\hline
       
       $r_{200\rm{c}} \, (\mathrm{Mpc)}$ & log-Flat & $\log_{10}(1.0/{\rm Mpc})$ & $\log_{10}(3.0/{\rm Mpc})$ \\
       $r_{\rho} \, (\mathrm{Mpc)}$ & log-Flat & $\log_{10}(0.2/{\rm Mpc})$ & $\log_{10}(5.0/{\rm Mpc})$ \\
       $r_\nu^{\rm H} \, (\mathrm{Mpc)}$ & log-Flat & $\log_{10}(1.4/{\rm Mpc})$ & $\log_{10}(2.1/{\rm Mpc})$ \\
       $\mathcal{A}_0$ & log-Flat & $\log_{10}0.5$ & $\log_{10}5.0$ \\
       $\mathcal{A}_\infty$ & log-Flat & $\log_{10}0.5$ & $\log_{10}5.0$\\
       $\alpha$ & Flat & 0.1 & 1.9 \\      
\hline
\end{tabular}
\tablefoot{In Sect. \ref{full_sample_mamposst}, $r_\rho=r_{\rm{s}}$, since NFW is used for the mass density profile. $\alpha$ is a free parameter only when the gNFW model is used instead of NFW (see Sect. \ref{subsec:syst}).} 
\label{tab:priors_flat}
\end{table}

We considered NFW and gOM as the reference models for mass and anisotropy profiles, respectively (this combination is favoured by the BIC; see Sect. \ref{subsec:syst}). The corresponding estimates for the parameters with mixed components are reported in the corresponding column of Table \ref{tab:diff_samples_results}. It is worth to point out that our constraints on $r_{\rm s}$ are consistent with the concentration values found by \citet{Gavazzi_2009}. Note that the best-fit values reported in every Table of this paper correspond to the maximum likelihood values. 
 
\begin{figure}
    \centering
    \includegraphics[width=0.4\textwidth]{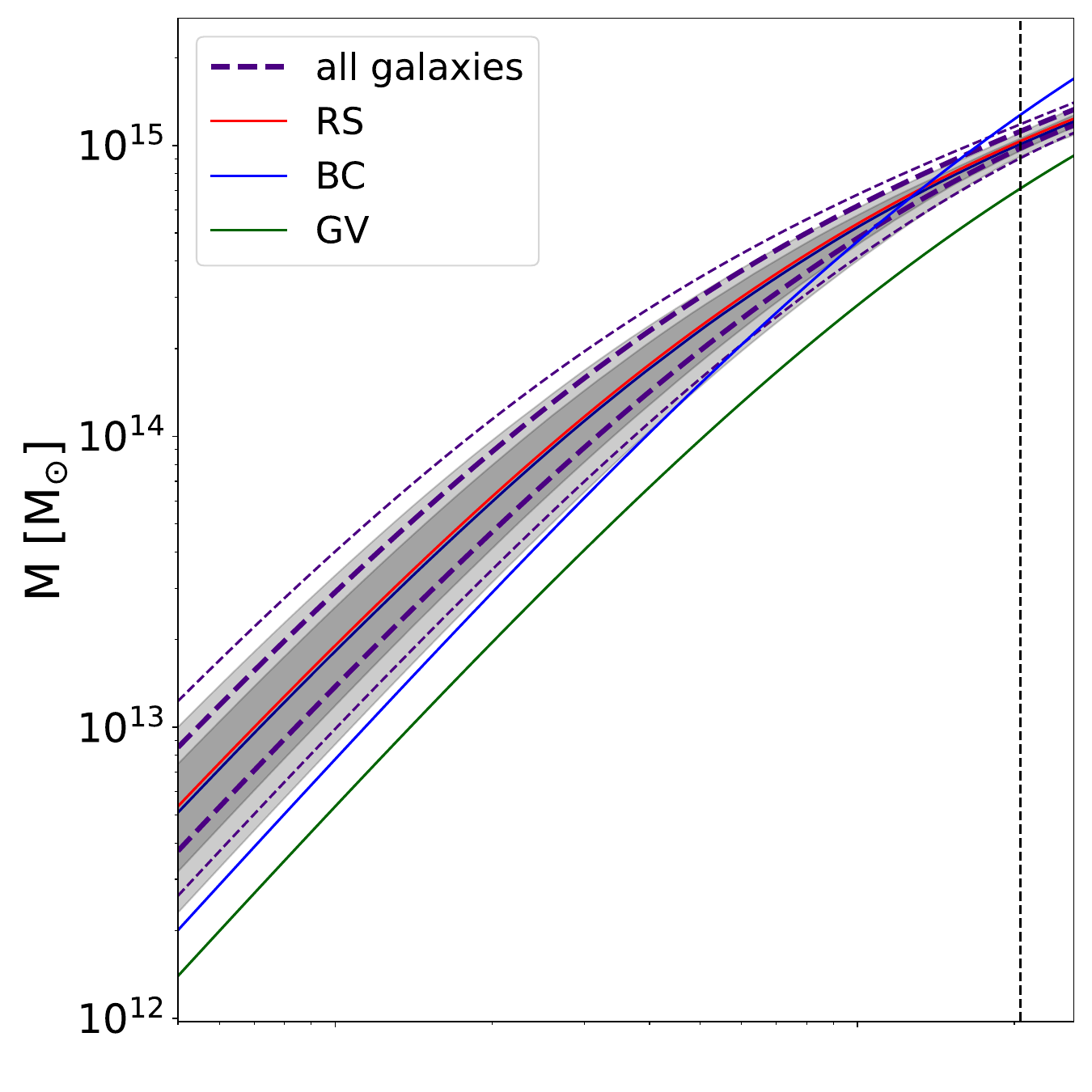}
    \includegraphics[width=0.40\textwidth]{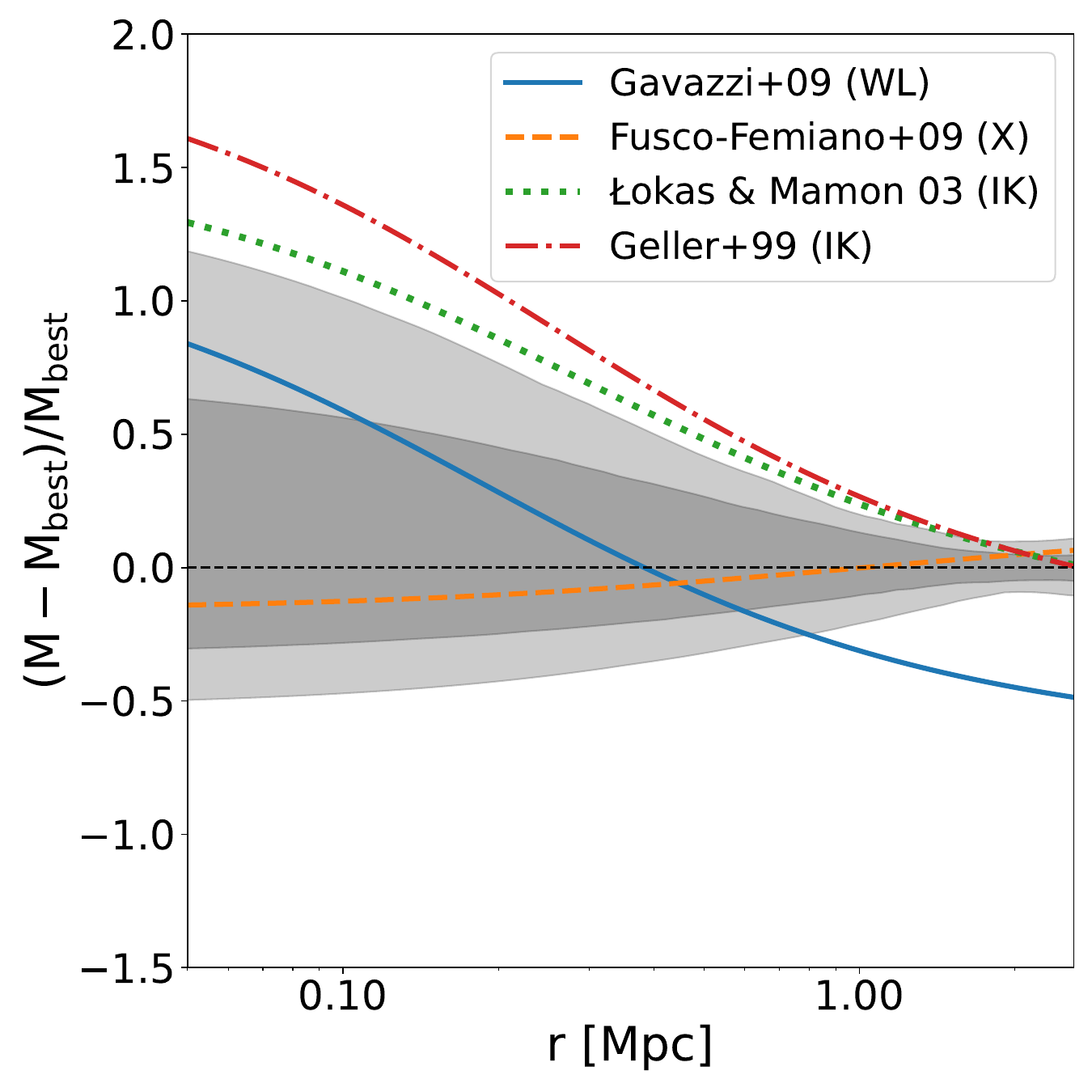}
    \caption{\textit{Top}: Estimated mass profiles in the \textsc{MG-MAMPOSSt} runs for the mixed components (1 and $2\,\sigma$ regions delimited by dashed purple lines), separated by colour classes (RS in red, GV in green, BC in blue) and produced by the joint likelihood (1 and $2\,\sigma$ regions coloured in dark and light grey, respectively, and vertical dashed line corresponding to the virial radius). \textit{Bottom}: Relative ratio between our best fit $M_{\rm best}(r)$ and other estimates of the full mass profile of Coma from the literature using different techniques, i.e. weak lensing (WL), X-ray (X) and internal kinematics (IK). Solid blue curve: \cite{Gavazzi_2009}. Dashed orange line: \cite*{Fusco2009}. Dotted green line: \cite{Lokas_2003}. Dash-dotted red curve: \cite*{Geller1999}. Dark and light grey shadings indicate 1 and $2\,\sigma$ confidence regions on our best fit, $M_{\rm best}(r)$. Confidence regions from the literature, $M(r)$, are not shown for the sake of clarity.}
    \label{fig:mass_profile}
\end{figure}

\begin{figure}
    \centering
    \includegraphics[width=0.9\columnwidth]{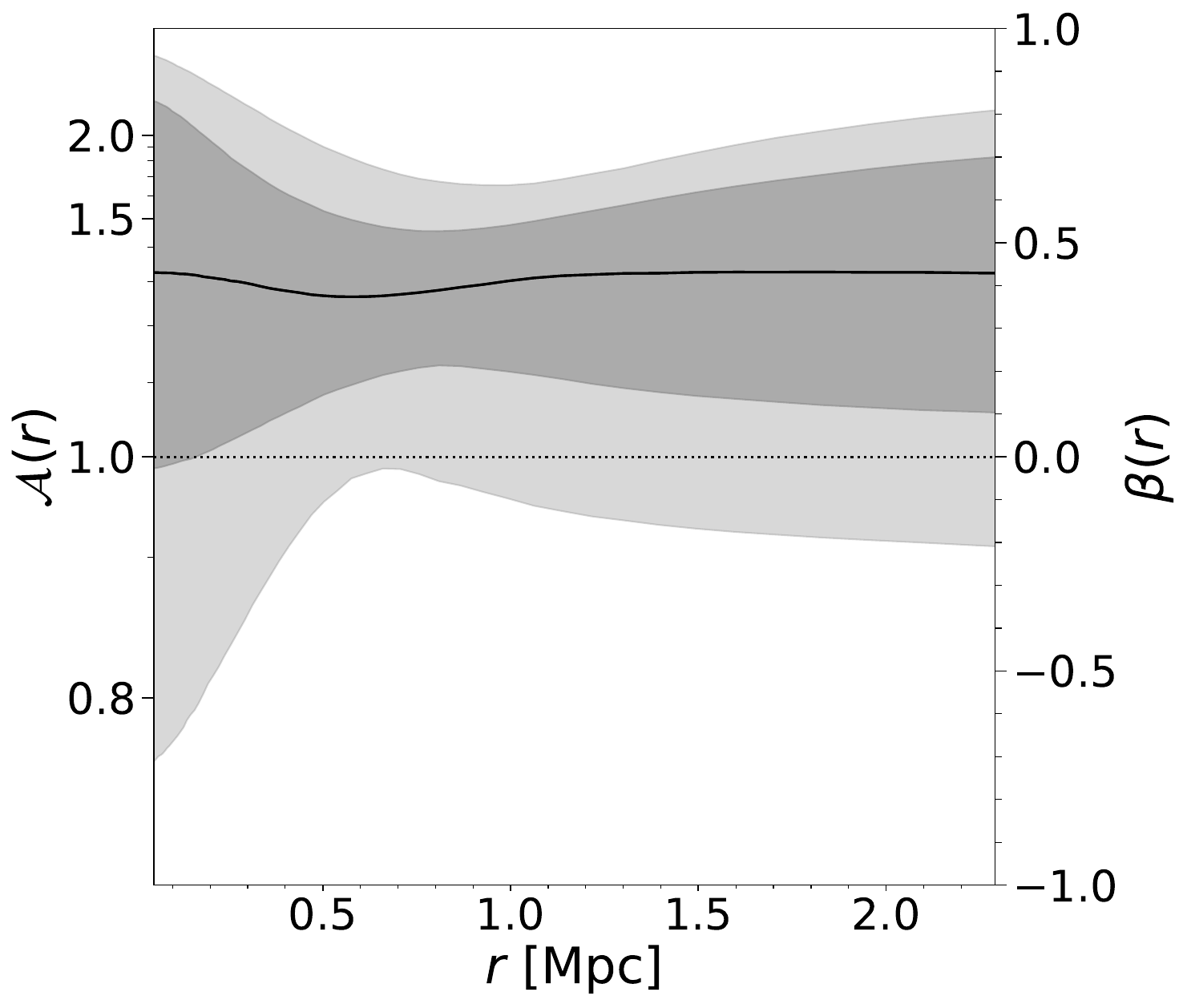}
    \caption{Estimated anisotropy profile in the \textsc{MG-MAMPOSSt} run for the mixed components with the NFW mass model and gOM anisotropy model. The solid curve is the median profile of the MCMC chain, while the horizontal dotted line highlights isotropic orbits.}
    \label{fig:anis_profile}
\end{figure}

For the mixed-population run, we plot the reconstructed mass profile along with its $1\,\sigma$ and $2\,\sigma$ uncertainty regions as the dashed and dotted purple lines, respectively, in Fig.~\ref{fig:mass_profile} (top panel).  The anisotropy profile, shown in Fig.~\ref{fig:anis_profile}, indicates the presence of radial orbits both in the inner and outer regions of the system, but still consistent with isotropy at $1\sigma$ and $2\sigma$ confidence levels, respectively. Moreover, the profile shows a negligible variation with radius.  
The prevalence of radial orbits reflects the presence of a population of recently accreted galaxies from the surrounding large-scale structure.

\medskip

We then ran \textsc{MG-MAMPOSSt} on the p.p.s. of the different colour classes according to the joint likelihood, defined as
\begin{equation}\label{eq:joint}
\begin{split}
    \ln{\mathcal{L}_\text{joint}} = & \ln{\mathcal{L}_\text{GV}}(r_{200\rm{c}},r_{\rm s},\vec{\theta}_{\rm GV}| D_{\rm GV} ) + \\
    &
    \ln{\mathcal{L}_\text{BC}}(r_{200\rm{c}},r_{\rm s},\vec{\theta}_{\rm BC}| D_{\rm BC} ) + \ln{\mathcal{L}_\text{RS}}(r_{200\rm{c}},r_{\rm s},\vec{\theta}_{\rm RS}| D_{\rm RS} )\,,
\end{split}
\end{equation}
where $\vec{\theta}_{i} = \{ \mathcal{A}_0, \mathcal{A}_\infty, r_{\nu}^{\rm H}\}_i$ is the vector of anisotropy and number density profile parameters for each population, $i = $ RS, BC, GV. $D_i$ indicates the dataset (p.p.s) used. The fit was then performed over the 11-dimensional parameter space $(r_{200 \rm{c}},r_{\rm s},\vec{\theta}_{i})$. Based on the analysis performed with the mixed components, here we kept the NFW and the gOM as the reference mass density and (color class-) anisotropy profile models. The employed priors were the same as those used for the mixed sample analysis (Table \ref{tab:priors_flat}), with the exception of $r_\nu^{\rm H}$, for which priors were adapted for each class following the results of Sect.~\ref{num_density_methods}. The number of galaxies between $0.10 \,\mathrm{Mpc}$ and $2.30 \,\mathrm{Mpc}$ -- the minimum and maximum radii considered for the reference analysis -- is 1030 for the RS, 129 for the GV, and 152 for the BC.  

Table \ref{tab:diff_samples_results} shows the MG-MAMPOSSt results. We find a mass normalisation of $r_{200\rm{c}}=2.07 \pm 0.05~({\rm stat})\pm0.07 ~({\rm syst}) \, \mathrm{Mpc}$, corresponding to $M_{200\rm{c}} = 1.04_{-0.08}^{+0.07}~({\rm stat}) \pm 0.09~({\rm syst}) \times 10^{15} \, \mathrm{M}_\odot $, and a scale radius for the mass profile of $r_{\rm s} =0.73_{-0.30}^{+0.24}\,({\rm stat}) \pm 0.21 \,({\rm syst)} \, \mathrm{Mpc}$. The systematic uncertainties take into account the variation in the best fit due to different choices of the mass and anisotropy models, as well as the variations in the projected radial range considered, cluster centre, and velocity rest frame (see Sect.~\ref{subsec:syst}). The mass profile and the corresponding $1$ and $2\,\sigma$ regions are shown as shaded grey areas in the top panel of Fig.~\ref{fig:mass_profile}. The bottom panel of the same figure displays the relative ratio between ours and other estimates from the literature for the total mass profile of the cluster. 

Overall, we found a general good agreement among different mass profiles within uncertainties at large radii, with the largest deviations found instead near the centre. The weak lensing analysis of \cite{Gavazzi_2009} predicts a cluster mass that is lower, but still in agreement with our value considering their $1\,\sigma$ uncertainties, as discussed in Sect.~\ref{sec:conclusions}. See also a comparison between our best virial mass measurement and those coming from more recent studies in Fig. \ref{fig:mass_compare}.

\begin{table}
\centering
\tabcolsep=2.2pt
\caption{Best-fit (maximum posterior) parameters for the runs with mixed and separated components.}
\renewcommand{\arraystretch}{1.3}
\begin{tabular}{lccccc}
\hline
Parameter & Mixed & RS & {GV} & {BC} & Joint \\
\hline
       
       $r_{200\rm{c}} \, (\mathrm{Mpc})$ & $2.14_{-0.05}^{+0.09}$ & $2.06_{-0.05}^{+0.07}$ & $1.57_{-0.18}^{+0.48}$ & $2.55_{-0.44}^{+0.21}$ & $2.07\pm 0.05$ \\
       $r_{\rm s} \, (\mathrm{Mpc})$ & $0.70_{-0.30}^{+0.22}$ & $0.61_{-0.37}^{+0.02}$ & $0.83_{-0.62}^{+0.85}$ & $2.42_{-2.18}^{+0.05}$ & $0.73_{-0.30}^{+0.24}$ \\
       $r_\nu^{\rm H} \, (\mathrm{Mpc})$ & $1.70_{-0.13}^{+0.11}$ & $1.27_{-0.08}^{+0.12}$ & $7.40_{-2.48}^{+2.60}$ & $5.17_{-1.72}^{+1.32}$ & / \\
       $\mathcal{A}_0$ & $1.10_{-0.33}^{+0.65}$ &
       $1.00_{-0.30}^{+0.68}$ & $0.81_{-0.25}^{+1.22}$ & $1.78_{-0.86}^{+0.79}$ & / \\
       $\mathcal{A}_\infty$ & $1.46_{-0.58}^{+0.20}$ & $1.60_{-0.58}^{+0.02}$ & $1.89_{-0.73}^{+1.44}$ & $0.99_{-0.29}^{+0.78}$ & / \\
          
\hline
\end{tabular}
\tablefoot{The employed mass and anisotropy models are NFW and gOM, respectively. The assumed priors, for these runs, are broad and flat, except for $r_\nu^{\rm H}$, for which we considered informative priors based on the best-fit values discussed in Sect. \ref{num_density_methods}. The three columns referring to the separated components report the results obtained with the single population runs.}
\label{tab:diff_samples_results}
\end{table}

We further performed runs on the single populations to infer the mass profile predicted by each color classes; the estimates of the free parameters, for each of the three independently treated populations, are reported in the corresponding columns of Table \ref{tab:diff_samples_results}. As expected, the kinematics are completely dominated by the RS galaxies, producing a profile in excellent agreement with that obtained from the mixed sample, as is further shown by the solid red curve in Fig.~\ref{fig:mass_profile} (top). While the analysis of the GV sample tends to underestimate the total mass at all radii, the BC galaxies favour a less concentrated and more massive cluster, supporting the idea that they are farther away from dynamical equilibrium.

Finally, we performed an additional run to estimate the diffuse dark matter mass profile of the cluster at $r> 0.10$ Mpc. We considered the mass profiles of the gas  \citep[estimated from the electron density profile in][]{Mirakhor_2020} and the stellar component, respectively, $M_\text{\rm gas}(r)$ and $M_{*}(r)$.
For the stellar mass, we obtained the parameters of a Hernquist surface density profile by fitting it to the stellar-mass-weighted distribution of projected radii.

\textsc{MG-MAMPOSSt} was then applied to fit a multi-component mass model as
\begin{equation}\label{eq:multic}
    M_\text{\rm tot}(r) = M_\text{\rm gas}(r) + M_{*}(r) + M_\text{\rm DM}(r)\,,
\end{equation}
where $M_\text{\rm DM}(r)$ is the DM distribution, parametrised as a gNFW model. The uncertainties on the stellar mass profile in galaxies and on the gas mass profile were accounted by first maximum-likelihood-fitting a Hernquist model and a $\beta$ model to the galaxies and gas data, respectively, and using the resulting covariance matrix as a multi-variate Gaussian input prior on the corresponding $M_{*}(r)$ and $M_\text{\rm gas}(r)$ in the \text{MG-MAMPOSSt} analysis.
Note that the gas mass uncertainties are not explicitly provided in the literature; as such, we have considered conservative uncertainties of $\sim 10\%$ in the parameters of the $\beta$ model as provided by the variation in the gas profile estimates of various works \citep[see e.g.][]{Mirakhor_2020, Churazov_2021}.

The resulting mass profiles are shown in Fig. \ref{fig:multicompoent}. The dark matter dominates at all radii, with a corresponding virial mass of $M_{200\rm{c}}^{\rm DM} = 8.6^{+1.2}_{-0.8} \times 10^{14} \, \text{M}_\odot$. The total mass at $r= 2.07$ Mpc, our best-fit value for the virial radius, is found to be $M_{\rm tot} = M_{\rm DM}+M_{\rm gas}+M_{*} = (1.06\pm{0.08})\times 10^{15} \,\text{M}_\odot$, in excellent agreement with the single-profile estimates. With this multi-component approach, we can further estimate the value of the baryon fraction as  $f_{\rm b} \equiv 1- M_{\rm DM}/M_{\rm tot} =  0.15 \pm 0.02$ at $r = 2.07$ Mpc, consistent with the cosmological value of $\Omega_{\rm b}/\Omega_{\rm m}$ (e.g. \citealt{Krolewski2025}). As for the inner slope of the DM profile, we obtain a value in agreement with what is found for the single-mass profile run (Sect. \ref{subsec:syst}), $\alpha = 1.80^{+0.08}_{-0.48}$, suggesting a steeper DM profile with respect to standard CDM predictions. 
When considering the physical parameter $r_{-2} = (2 - \alpha)\,r_{\rm s}$, we obtain a constraint of $r_{-2} = 0.62^{+0.38}_{-0.30}$, which is in excellent agreement with the value of $r_{\nu,-2} = 0.64^{+0.02}_{-0.04}$ of the scale radius of the RS galaxies. This confirms that passive red galaxies tend to follow the shape of the DM distribution.

\subsection{Robustness of mass-orbital modelling}
\label{subsec:syst}

The \textsc{MG-MAMPOSSt} procedure of modelling the mass and orbital profiles of complex systems such as the Coma cluster produces results that inevitably depend on several choices and assumptions, whose validity and robustness must therefore be assessed. First of all, one has to assemble the input p.p.s. sample of member galaxies, which, even after the removal of interlopers, is dependent on two main variables:

\begin{enumerate}
    \item[\textit{(i)}] The co-ordinate centre of the cluster, which we located at (RA=194.90 deg, DEC = 27.96 deg), the position of NGC 4874.
    
    \item[\textit{(ii)}] The LoS velocity of the rest frame of the cluster, chosen as the biweight mean LoS velocity of its members. 
    
\end{enumerate}    
The input parametrisations of the mass (\textit{iii}) and anisotropy profiles (\textit{iv}) (as well as of the galaxy number density profile, see Sect. \ref{num_density_methods} and Appendix \ref{app:num_density_assump}), with the choice of their relative priors, introduce additional potential sources of uncertainty.

\begin{figure}
    \centering
    \includegraphics[width=0.9\linewidth]{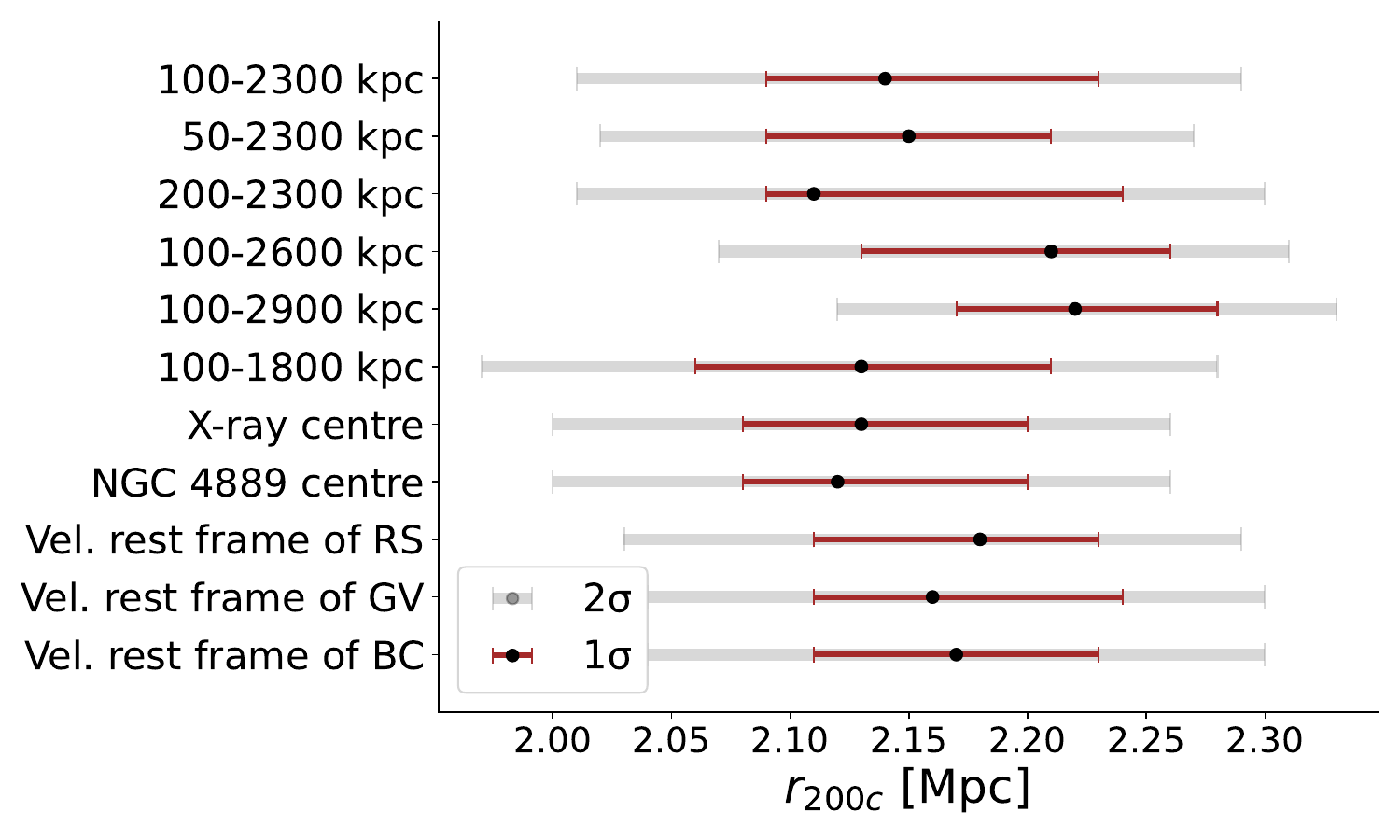}
    \includegraphics[width=0.9\linewidth]{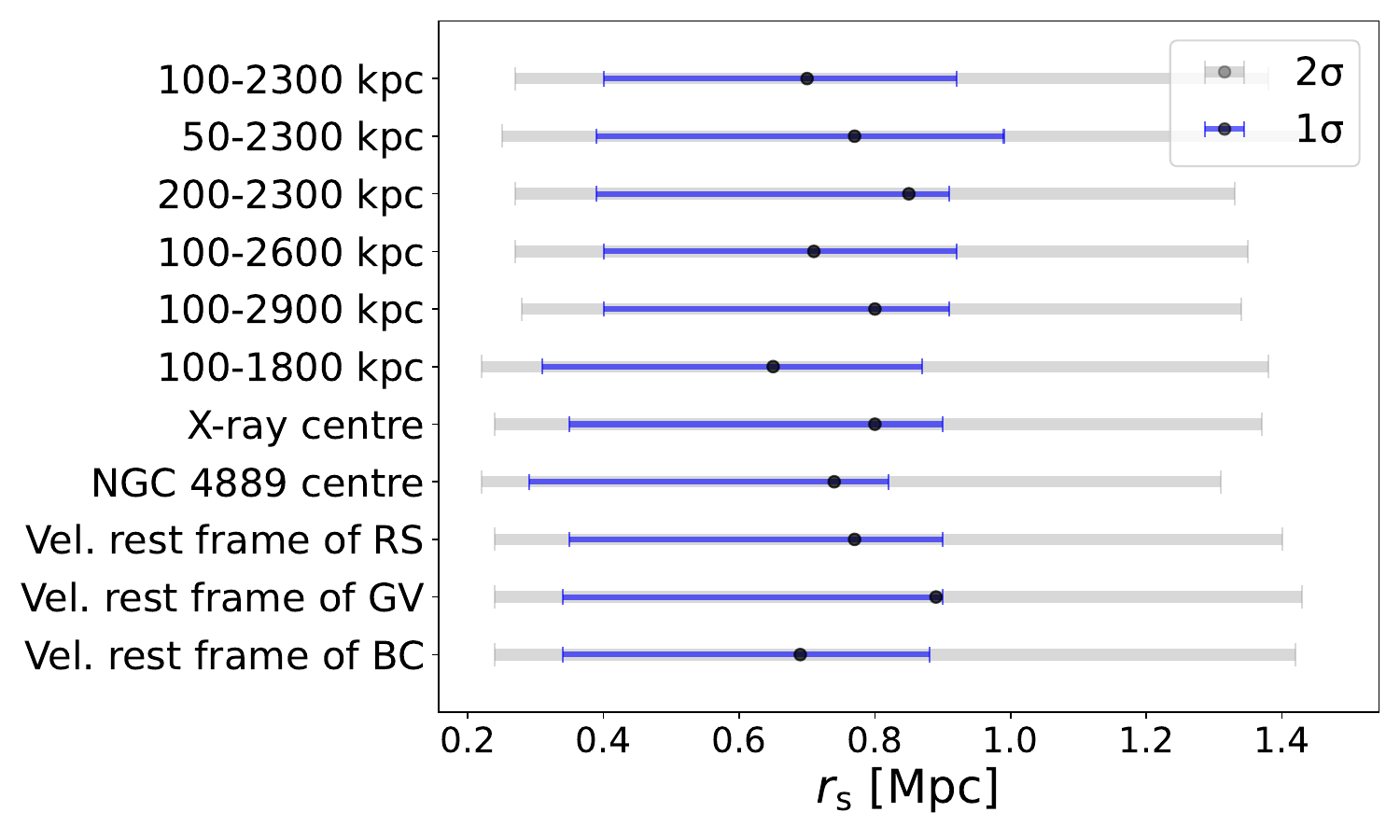}
    \caption{$r_{200\rm{c}}$ (\textit{top}) and $r_{\text{s}}$ (\textit{bottom}) estimates and confidence intervals ($1$ and $2\,\sigma$) for the different runs, specified on the left, performed to check the robustness of the mass modelling. The top line in each panel represents the reference \textsc{MG-MAMPOSSt} run.}
    \label{fig:r200-rs-comparison}
\end{figure}

The projected radial range within which galaxies are considered in the fit plays a fundamental role. This depends on:

\begin{enumerate}
    \item[\textit{(v)}] The minimum projected radius, which we chose to be 100 kpc, in order to exclude the innermost region where dynamics is dominated by NGC 4874 and NGC 4889.
    
    \item[\textit{(vi)}] The maximum projected radius, which we chose to be 2.30 Mpc, in order to exclude the outer regions, where we cannot ensure dynamical equilibrium. 
    
\end{enumerate} 
In this section we want to verify that reasonable variations in the choices of \textit{i)--vi)} lead to results that are within the uncertainties of our reference analysis of Sect. \ref{full_sample_mamposst}. We therefore re-ran \textsc{MG-MAMPOSSt} to perform the following tests:

\begin{enumerate}

    \item[\textit{(i)}] We moved the co-ordinate centre of the cluster to the position of NGC 4889 (RA=195.03 deg, DEC=27.98 deg) and to the position of the X-ray peak of emission (RA=194.94 deg, DEC=27.95 deg; \citealt{Sato_2011}).
    
    \item[\textit{(ii)}] We shifted the LoS velocity of the rest frame of the cluster to the average LoS velocity of RS (which is lower by $\sim 20$ km/s), GV, or BC galaxies (which is higher by $\sim 53$ and $\sim 4$ km/s, respectively). 

    \item[\textit{(iii)}] We changed the assumed mass density model from NFW to Hernquist, Burkert, and gNFW.

    \item[\textit{(iv)}] We changed the assumed anisotropy model from gOM to gT.

    \item[\textit{(v)}] We changed the minimum projected radius to both 50 kpc and 200 kpc.
    
    \item[\textit{(vi)}] We changed the maximum projected radius to 1.8 Mpc, 2.6 Mpc, and 2.9 Mpc.
    
\end{enumerate} 
These tests have been run on the mixed sample of galaxies, since the current implementation of the joint procedure is not yet optimised. A full re-run of all configurations would therefore be prohibitively expensive in terms of computing time, and we expect the same systematic effects to apply on both methodologies. We used the same priors of Table \ref{tab:priors_flat} to not introduce other differences that could invalidate the tests. The comparison of the best-fit estimates of $r_{200\rm{c}}$ and $r_{\rm s}$ is shown in Fig.~\ref{fig:r200-rs-comparison}. All the results are fully in agreement with each other; the largest differences for $r_{200\rm{c}}$ are observed when changing the maximum radius considered, producing an increase of $\sim 5\%$. This effect can probably be ascribed to the fact that the assumption of dynamical equilibrium is not valid up to large radii such as 2.6 and 2.9 Mpc, causing \textsc{MG-MAMPOSSt} to overestimate $r_{200\rm{c}}$; the decrease in spectroscopic completeness at such large radii (see Appendix \ref{app:completeness}) is not relevant for the results, as we proved by fitting the galaxy number density profile, assigning each tracer a weight based on the completeness profile and finding negligible variations in $r_\nu^{\rm{H}}$.

The biggest differences for $r_{\rm s}$ arise when shifting to the GV rest frame, causing a variation of $\sim 30\%$ in the best fit, still well within the much larger uncertainties of $r_{\rm s}$ with respect to $r_{200\rm{c}}$. Different combinations of $M_{\rm dyn}(r)$ and $\beta(r)$ models provided mass estimates in agreement within the uncertainties, with smaller variations with respect to those reported in Fig. \ref{fig:r200-rs-comparison}. The resulting BIC values suggest that the NFW model for the mass profile is favoured with respect to the Burkert and Hernquist models, and the gOM anisotropy model is slightly better than the gT model, but differences in the Bayesian evidence values are negligible. When running with the gNFW, the inner slope of the mass density profile is estimated as $\alpha=1.71_{-0.39}^{+0.16}$ \citep[similarly to what found in the 94 stacked regular WINGS clusters in ][]{Mamon_2019}, at the cost of the mass profile having a totally unconstrained scale radius. The BIC for this run is 7 units higher (worse) than the reference run, but the difference reduces to only 2 units when using the AIC criterion, defined as

\begin{equation}
    \text{AIC} = 2\,k \, - 2 \, \ln\mathcal{L} \, ,
    \label{eq:AIC}
\end{equation}
which, in our case, reduces the penalty caused by the increasing of the number of free parameters.

\begin{figure}
    \centering
    \includegraphics[width=0.9\linewidth]{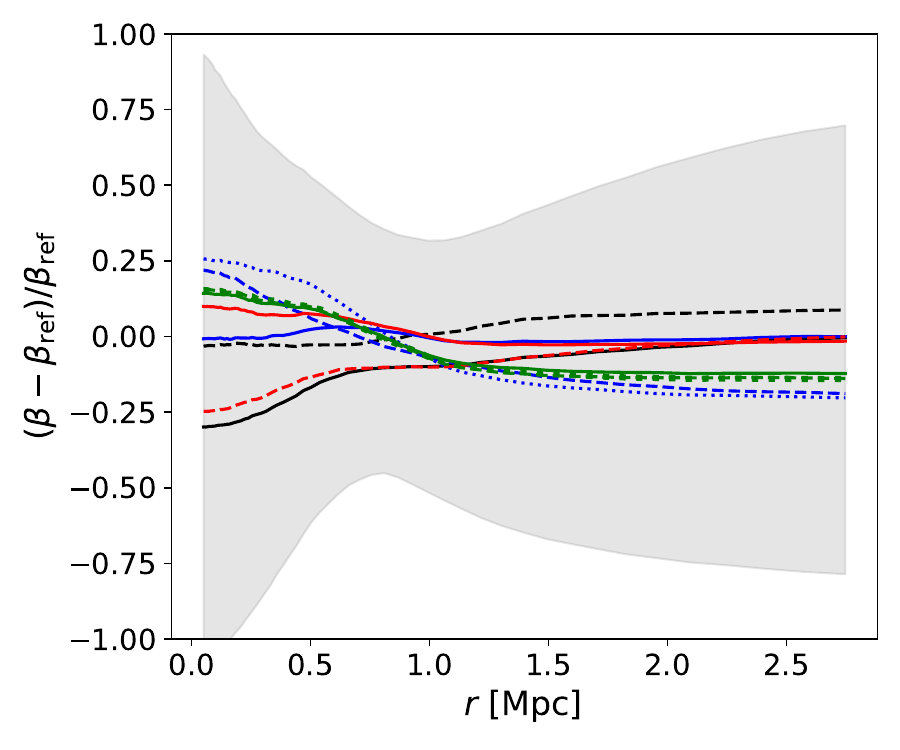}
    \caption{Relative variation in the anisotropy profiles with respect to the reference profile ($\beta_{\text{ref}}$) of Fig.~\ref{fig:anis_profile} after: changing minimum radius (black) to 50 kpc (solid) and 200 kpc (dashed); changing maximum radius (blue) to 1.8 Mpc (solid), 2.6 Mpc (dashed) and 2.9 Mpc (dotted); moving the cluster centre (red) to the co-ordinates of NGC 4889 (solid) and of the X-ray peak (dashed); shifting the velocity rest frame (green) to that of RS (solid), BC (dashed) and GV (dotted) galaxies. The grey band identifies the $1\,\sigma$ uncertainty band of the reference profile.       }
    \label{fig:relative_anis}
\end{figure}

Figure~\ref{fig:relative_anis} shows the relative variation in the anisotropy profiles produced by the runs described in this section with respect to the reference profile of Fig.~\ref{fig:anis_profile} (we exclude those obtained from (\textit{iii}) and (\textit{iv}) for visual clarity): all the profiles are well within the $1\,\sigma$ uncertainty regions of the reference run. We therefore conclude that the mass-orbital modelling of Coma with \textsc{MG-MAMPOSSt} is very robust and only slightly affected by systematics.

\subsection{Orbital anisotropy study for different colour classes}
\label{subsec:diff_sample_mamposst}

We now compare the anisotropy profiles of RS, GV, and BC galaxies. In fact, even if the profile obtained with the mixed sample does not show significant trends with radius (Fig.~\ref{fig:anis_profile}), it might be that different populations contribute differently to the orbital properties of the cluster. 

In general, given the evolutionary picture of galaxies in clusters described in Sect. \ref{sec:intro}, we would expect to observe a higher value of the anisotropy for BC and GV with respect to RS galaxies. However, the runs performed in Sect. \ref{full_sample_mamposst} with single colour classes cannot be used to directly compare the anisotropy profiles, because the corresponding mass parameters are significantly different. While the values of $r_{\rm s}$ obtained from the different tracers are mutually consistent (note that the BC estimate presents very high uncertainties), this is not the case for $r_{200\rm{c}}$, for which the GV estimate is lower and the BC one is higher than that of the RS. This BC result supports the hypothesis proposed in Sect. \ref{subsec:LoS_methods}; namely, that a substantial fraction of BC galaxies belongs to a filamentary-like structure along the LoS, leading to an overestimation of the cluster's virial radius. 

\begin{figure*}
    \centering
    \includegraphics[width=0.2815\linewidth]{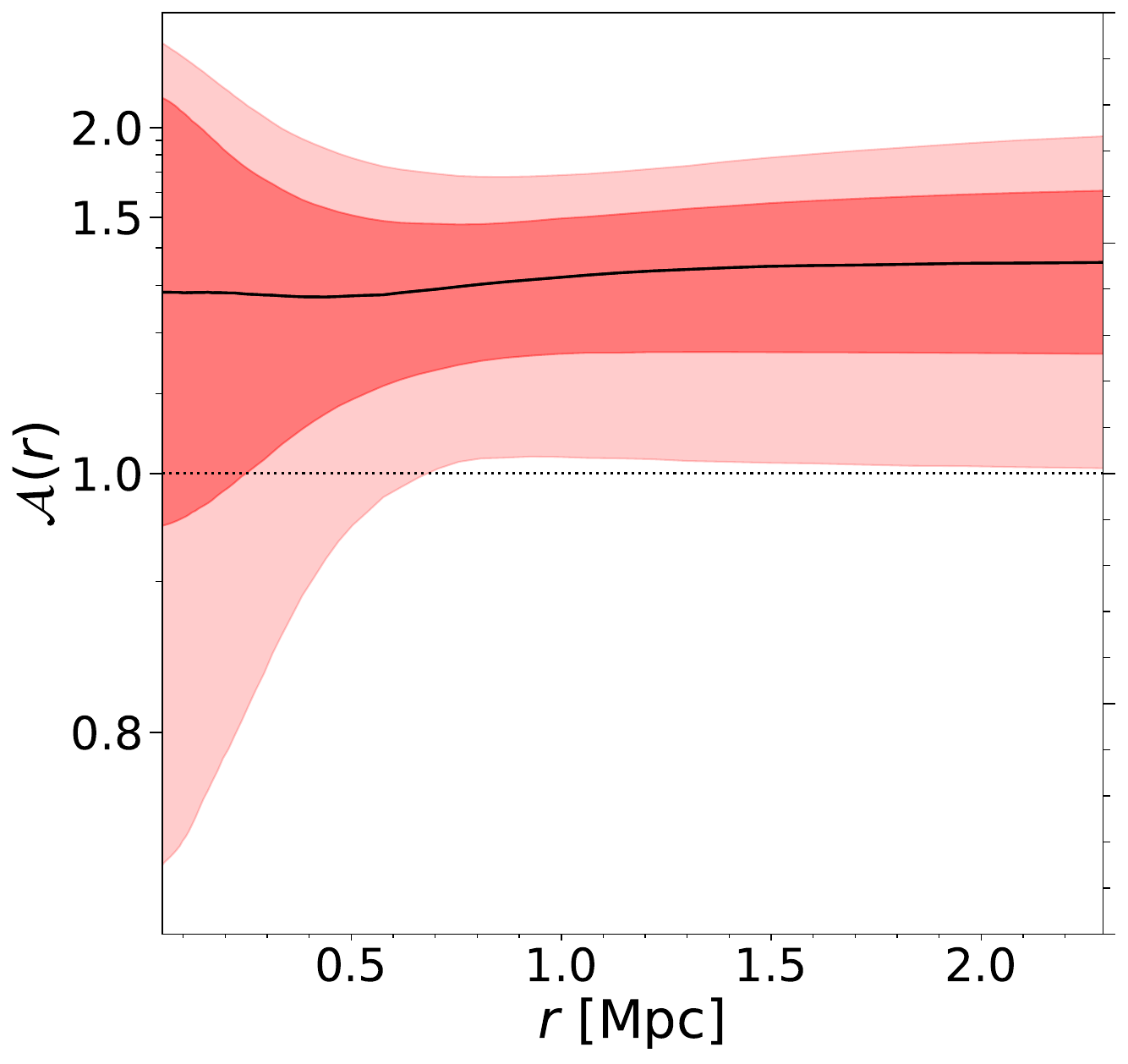}
    \includegraphics[width=0.244\linewidth]{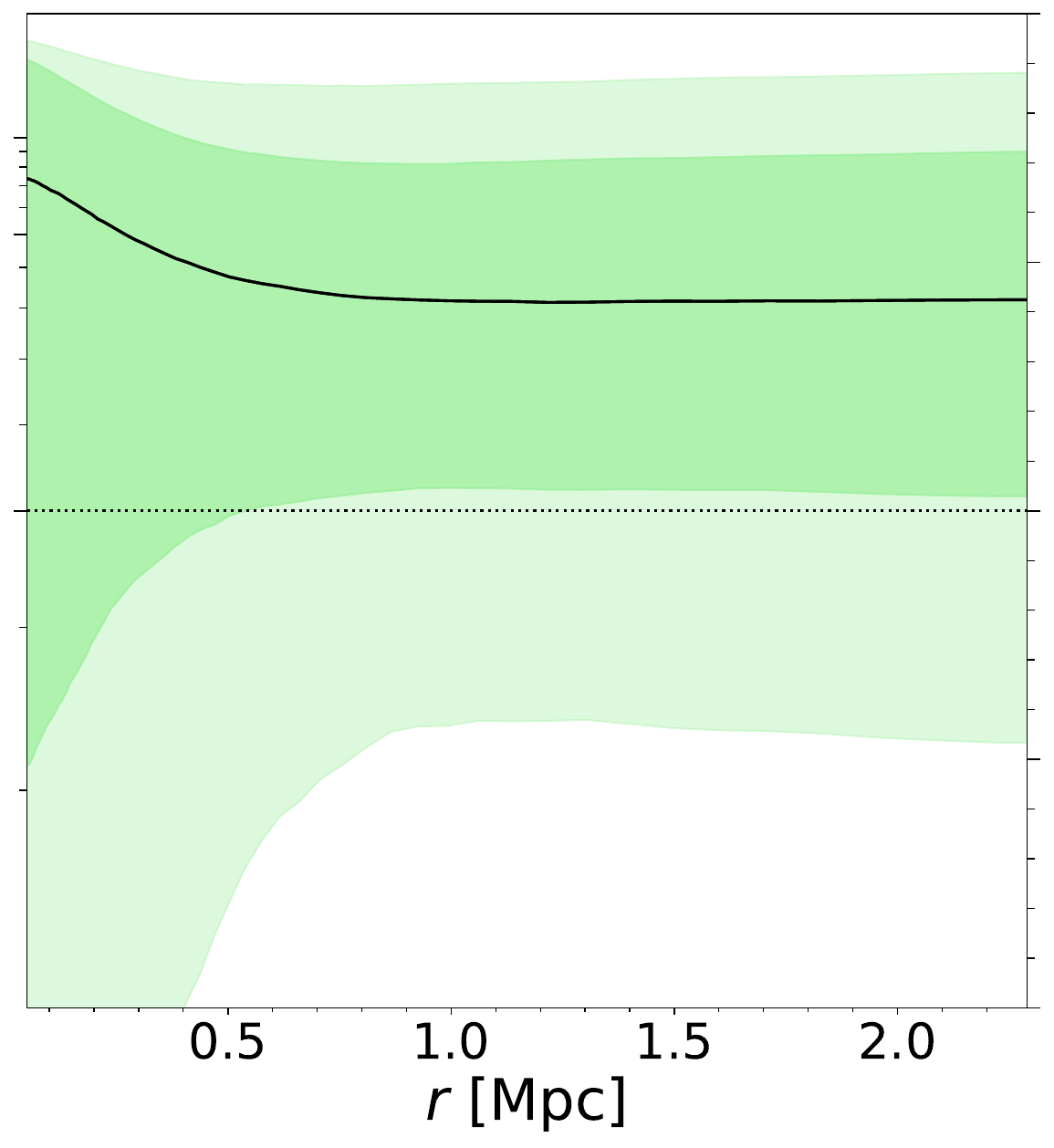}
    \includegraphics[width=0.2855\linewidth]{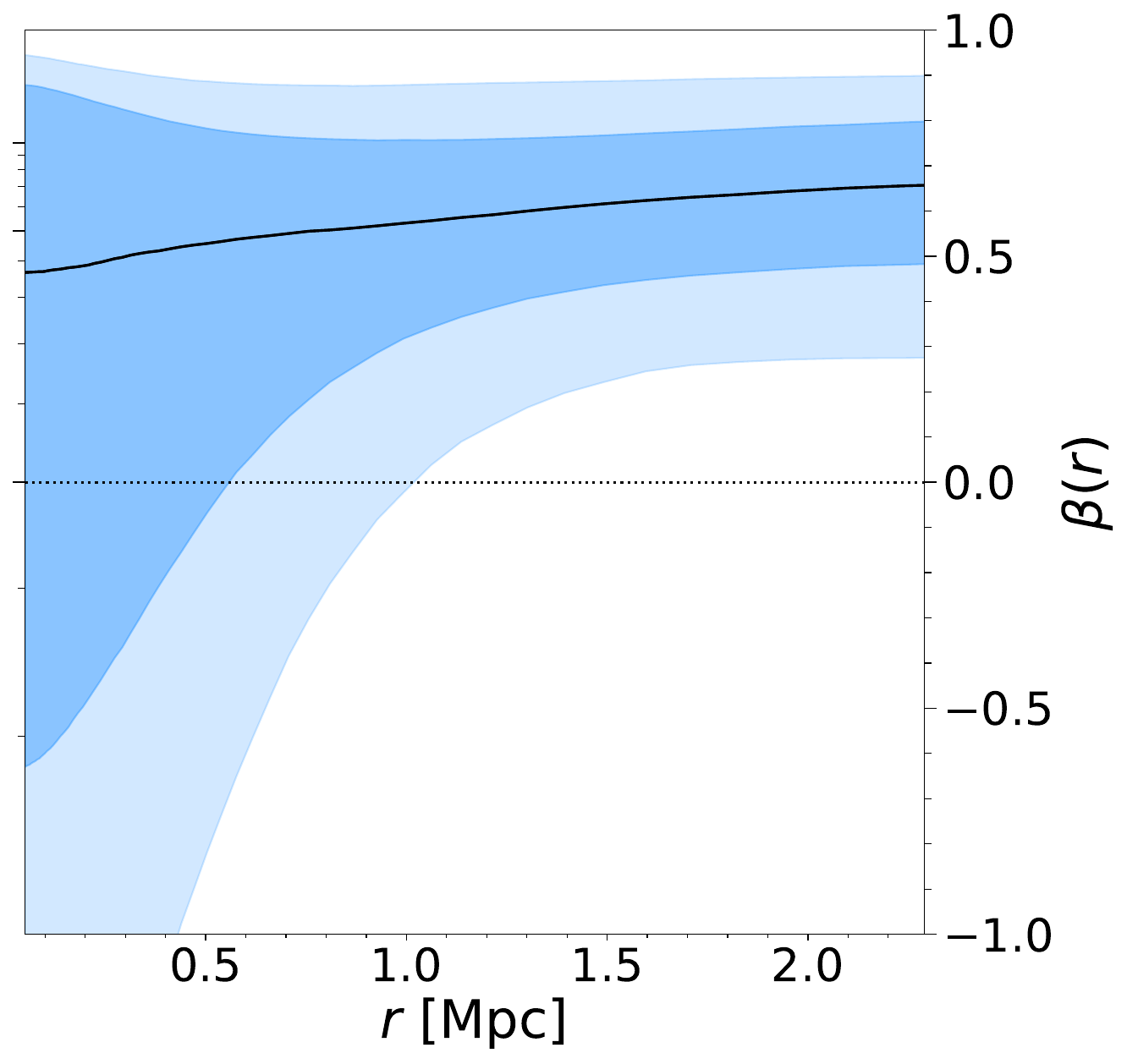}
    \caption{Estimated anisotropy profiles in NFW gOM population-joint run of Sect. \ref{full_sample_mamposst} for RS (\textit{left}), GV (\textit{middle}), and BC (\textit{right}) galaxies. The coloured and shaded regions indicate 1$\,\sigma$ and 2$\,\sigma$ confidence intervals, respectively. Solid curves are the median profiles of the MCMC chain, while horizontal dotted lines highlight isotropic orbits. }
    \label{fig:anis_subsamples}
\end{figure*}

In order to directly compare the anisotropy profiles of the populations, we needed their mass parameters to be consistent a priori; therefore, we considered the colour-classes anisotropy profiles provided by the population-joint likelihood of Eq. \ref{eq:joint}. The corresponding best-fit estimates for $\vec{\theta}_i$, with $i=$RS, GV, and BC, are reported in Table \ref{tab:diff_samples_results_withpriors}.
\begin{table}
\centering
\caption{Best-fit (maximum posterior) estimates of number density scale radii and anisotropy parameters from the joint run of Sect. \ref{full_sample_mamposst}}
\renewcommand{\arraystretch}{1.3}
\begin{tabular}{lccc}
\hline
{Parameter} & {RS} & {GV} & {BC} \\
\hline
       
       $r_\nu^{\rm H} \, (\mathrm{Mpc})$ & $1.27_{-0.08}^{+0.10}$ & $7.78_{-3.48}^{+0.75}$ & $6.62_{-1.52}^{+1.76}$ \\
       $\mathcal{A}_0$ & 
       $1.20_{-0.46}^{+0.47}$ & $1.76_{-0.95}^{+1.71}$ & $1.37_{-0.40}^{+1.08}$ \\
       $\mathcal{A}_\infty$ & $1.50_{-0.40}^{+0.07}$ & $1.24_{-0.38}^{+0.40}$ & $1.62_{-0.40}^{+0.60}$ \\
          
\hline
\end{tabular}
\tablefoot{Flat informative priors on $r_\nu^{\rm H}$ are based on the estimates of Sect. \ref{num_density_methods}, while those for $\mathcal{A}_0$ and $\mathcal{A}_\infty$ are flat and uninformative (same as Table \ref{tab:priors_flat}).}
\label{tab:diff_samples_results_withpriors}
\end{table}
The obtained anisotropy profiles of the different populations are shown for RS, GV, and BC galaxies in Fig.~\ref{fig:anis_subsamples}. By also comparing these plots with Fig.~\ref{fig:anis_profile}, relative to the profile obtained with the mixed sample, we note that all the profiles are similar within the uncertainties. The orbits are predominantly radial in the outskirts and in the centre, where the broader 1 and $2\,\sigma$ regions (within $R \lesssim 1.0 \, \mathrm{Mpc}$) provide consistency with isotropy. However, some differences in the median profiles are observed: in particular, the behaviour of RS galaxies is very similar to that of the mixed sample, showing a general tendency towards radial orbits both in the central regions and in the outskirts. The BC galaxies, on the other hand, display even more radial orbits in the outskirts, while remaining similar to the RS population in the core. Finally, GV galaxies exhibit an increase in radial anisotropy in the central region (where, however, uncertainties are quite high), whereas their outer orbits remain comparable to those of RS galaxies. These results are consistent with a scenario in which young, star-forming galaxies fall into the cluster from the outer regions and transition to the GV phase as they approach the cluster core, likely as a consequence of substantial ram-pressure stripping. 

This evolutionary pathway could be investigated by splitting the sample in other subclasses (e.g. from the star formation rate (SFR) vs $M_*$ diagram) but we do not expect significant differences in the results. As a simple test, we split the full sample of galaxies into two equal population stellar mass bins and ran \textsc{MG-MAMPOSSt} separately for the two classes, first assuming broad and flat priors to estimate the mass parameters and then employing Gaussian priors on $r_{200\rm{c}}$ and $r_{\rm s}$, based on the colour-joint likelihood estimates, to compare the anisotropy profiles. Table \ref{tab:massbins_results} indicates that low-mass galaxies estimate a slightly higher cluster virial mass and scale radius (even if consistent within the uncertainties), as expected given the larger fraction of BC galaxies, while both populations show a similar predominantly radial anisotropy. The details are reported in Appendix \ref{app:high-vs-low-mass}.

\section{Conclusions and discussion}
\label{sec:conclusions}

In this work, we performed a detailed kinematic analysis of the Coma cluster using a novel rich spectroscopic catalogue of cluster members from the DESI survey. We applied the \textsc{MG-MAMPOSSt} code to a sample of 1324 member galaxies in the projected phase space ($R, v_\text{RF}$) up to $R \sim 2.3$ Mpc, in order to jointly reconstruct the mass and orbital anisotropy profiles of Coma. By working with different subsamples of member galaxies, selected based on their colour, we obtained useful insights about the effect of the internal cluster dynamics on the properties of member galaxies.

\begin{figure}
    \centering
    \includegraphics[width=1\linewidth]{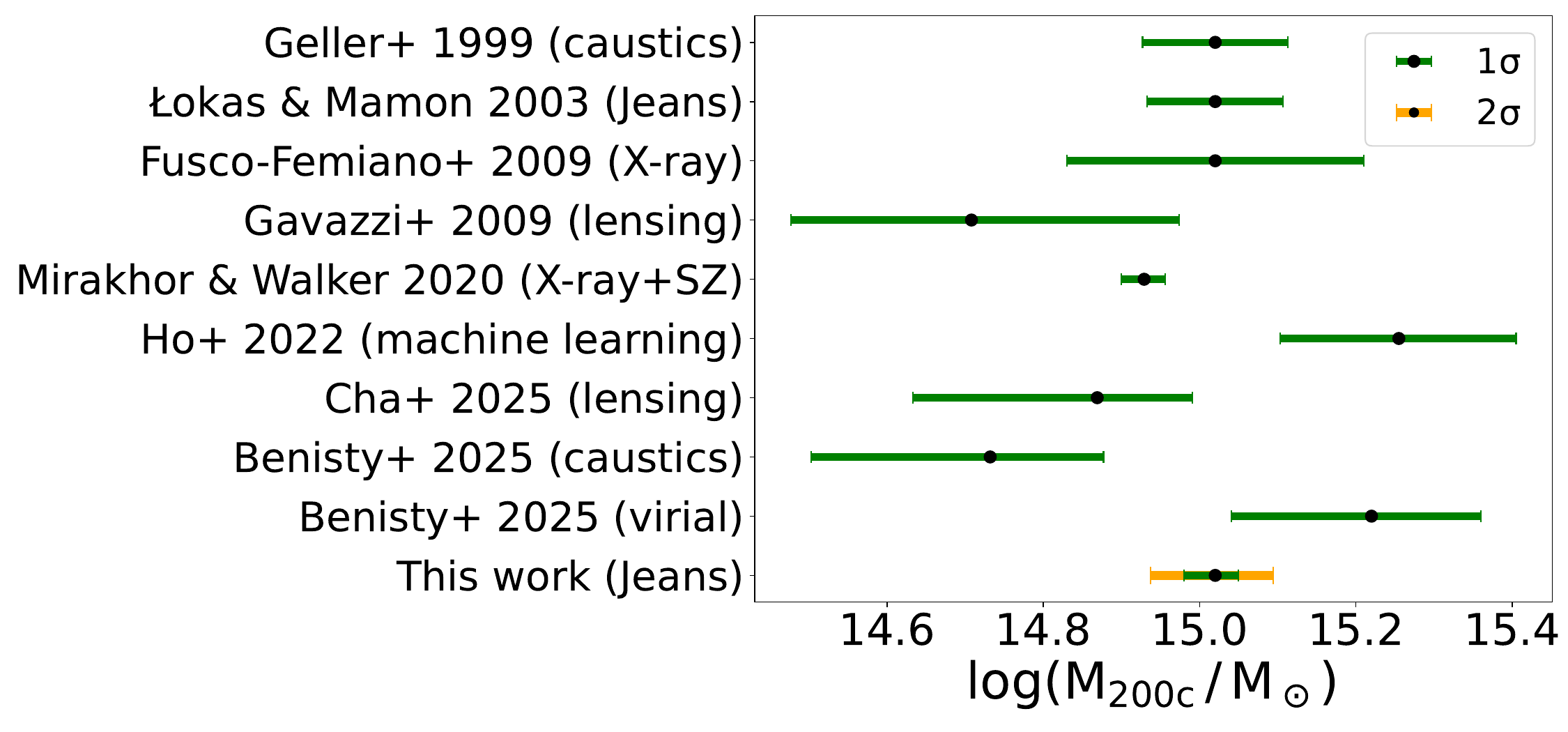}
    \caption{Comparison between $M_{200\rm{c}}$ estimates from different studies (indicated on the left of the figure, together with the used technique) and our best estimate from \textsc{MG-MAMPOSSt} with DESI dataset. From top to bottom: \citet{Geller1999}, \citet{Lokas_2003}, \citet{Fusco2009}, \citet{Gavazzi_2009}, \citet{Mirakhor_2020}, \citet{Ho_2022}, \citet{Cha_2025}, and \citet{Benisty25}.}
    \label{fig:mass_compare}
\end{figure} 

We found that BC galaxies, contrary to RS galaxies, exhibit strong deviations from Gaussianity in their LoS velocity distribution. This behaviour is consistent with theoretical predictions about LTGs still infalling towards the cluster centre, along more elongated orbits, in which the efficiency of ram-pressure stripping is higher in quenching star formation, thereby contributing to their eventual transition into ETGs. 

We ran \textsc{MG-MAMPOSSt} on the full mixed sample of member galaxies and on different colour subsamples, exploring various combinations of mass and anisotropy models. While all the models provide adequate fits of the total mass distribution, the NFW profile with gOM anisotropy gives the best Bayesian evidence; we quote the Coma cluster's virial mass, obtained from the joint likelihood of the different populations, to be $M_{200\rm{c}} = 1.04_{-0.08}^{+0.07}~({\rm stat}) \pm 0.09~({\rm syst}) \times 10^{15}  \, \mathrm{M_\odot}$, corresponding to $r_{200\rm{c}}=2.07 \pm 0.05~({\rm stat})\pm0.07~({\rm syst}) \, \mathrm{Mpc}$, where the systematic uncertainties account for the variation in the mode of the distribution across the different combinations of models and configurations tested. For the scale radius of the mass profile, we found $r_{\rm s}=0.73_{-0.30}^{+0.24}\,({\rm stat}) \pm 0.21 \,({\rm syst)} \, \mathrm{Mpc}$. When separately including the gas and stellar mass in galaxies in the fit, we estimate a dark matter virial mass  $M_{200\rm{c}}^{\rm DM} = 8.6^{+1.2}_{-0.8}\times 10^{14}\, \,\text{M}_\odot$ and a baryon fraction $0.15 \pm 0.02$ at $r = 2.07$ Mpc, in agreement with current cosmological studies.

Figure~\ref{fig:mass_compare} shows a graphic comparison between our $M_{200\rm{c}}$ estimate and those from other studies, in which different techniques have been adopted: weak lensing analysis, combination of X-ray and SZ effect, machine learning, caustics, virial theorem methods, and Jeans analysis. Interestingly, while our best-fit value is in excellent agreement with older studies (as is also shown in the bottom panel of Fig. \ref{fig:mass_profile}), it lies above the masses more recently derived from lensing \citep{Gavazzi_2009, Cha_2025}, X-ray+SZ \citep{Mirakhor_2020}, and caustics methods \citep{Benisty25}, but below the ones inferred from machine-learning \citep{Ho_2022} and virial theorem methods \citep{Benisty25}. The $2\,\sigma$ error bars of our measurement overlap with the $1\,\sigma$ error bars of lensing, X-ray+SZ, and virial theorem estimates, while approaching, but not reaching, the $1\,\sigma$ error bars from machine learning and caustics. Finally, our estimate is in excellent agreement with the halo mass inferred for Coma from its counterpart in the Manticore-Local simulations, using Bayesian field-level inference \citep{McAlpine_2025}.

The anisotropy profile shows a tendency towards radial orbits both in the centre and in the outskirts, similarly to the profile obtained by \citet{Pizzuti_2025b} for the cluster PSZ2 G067.17+67.46. The average $\beta(r)$ found by \citet{Biviano25} analysing nine massive clusters from the CLASH-VLT dataset shows a similar trend, with a slightly more isotropic behaviour in the centre, as well as the orbits found for ram-pressure-stripped member galaxies in nearby clusters \citep{Biviano24}. The stacking of 94 mock clusters by \citet{Mamon_2010} produced a very similar radial behaviour around the virial radius, while again predicting orbits closer to isotropy in the centre. A similar anisotropy profile was also obtained by \citet{Abdullah_2025} for simulated Coma-like clusters ($z=0$ and $\log(M/M_\odot) \geq 14.9$), which show a general radial tendency in the virial region (their analysis extends to much larger radii than ours), increasing from $\beta \sim 0.2$ in the centre to $\beta \sim 0.5$ around $r_{200\rm{c}}$ (see their fig. 3). 

The parameter's behaviour remains comparable between RS, GV, and BC galaxies, showing mainly radial orbits in the outskirts and in the centre, where broader uncertainties provide consistency with isotropy; however, the median profiles of GV and BC galaxies show even more substantial radial tendencies in the centre and in the outskirts, respectively, while RS galaxies produce a similar profile to the mixed sample. These results are consistent with many previous anisotropy studies of different galaxy populations, which report no significant differences between the orbits of passive and star-forming galaxies in clusters \citep[e.g.][]{Hwang_2008, Biviano_2013}.
The more radial behaviour of blue galaxies in the outer regions with respect to the red ones was also seen by \citet{Mamon_2019} in their analysis of stacked WINGS clusters (see their fig. 7).

Also, our virial mass estimates significantly differ depending on the employed tracers, with BC galaxies producing a much larger value; this may be an indication of a filamentary-like structure along the LoS, which we shall thoroughly investigate in the upcoming paper II. The new study will focus on the substructure content of the Coma cluster, the possibility of a recent major merger (see \citealt{Biviano+96}), and the implications of the dynamical state in the mass-orbit modelling.

In summary, this work provides an important example of how studying galaxy clusters properties through a kinematic approach is a very useful strategy to constrain different scenarios of galaxy evolution. With the advent of new wide-field spectroscopic surveys, providing larger and more complete samples of galaxies in clusters, this approach will enable increasingly detailed and statistically robust studies of cluster dynamics and galaxy accretion, thereby offering new insights into the role of the environment in shaping galaxy properties.

\section*{Data availability}
\label{sec:data_availability}

Table 4 is only available in electronic form at the CDS via anonymous ftp
to cdsarc.u-strasbg.fr (130.79.128.5) or via
\url{http://cdsweb.u-strasbg.fr/cgi-bin/qcat?J/A+A/}. 
It contains the DESI galaxy identifiers, equatorial co-ordinates, redshifts, and $g$- and $r$-band magnitudes of the initial sample of 2087 galaxies prior to any magnitude cut or cluster membership selection.

\begin{acknowledgements}
LP acknowledges support by the Italian Ministry for Research and University (MUR) under 457 Grant ``Progetto Dipartimenti di Eccellenza 2023-2027'' (BiCoQ).
LP and AR acknowledge the usage of INAF-OATs IT framework~\citep{Taffoni2020,Bertocco2020}.
AR acknowledges EuroHPC Joint Undertaking for awarding the project ID EHPC-REG-2024R01-029  access to Leonardo at CINECA, Italy.
AR acknowledges ISCRA for awarding this project access to the LEONARDO supercomputer, owned by the EuroHPC Joint Undertaking, hosted by CINECA, Italy (HP10BUFI59).
\end{acknowledgements}

\clearpage

\bibliographystyle{aa}
\bibliography{biblio}

\appendix
\section{Spectroscopic completeness}
\label{app:completeness}

\begin{figure}[!h]
    \centering
    \includegraphics[width=0.9\linewidth]{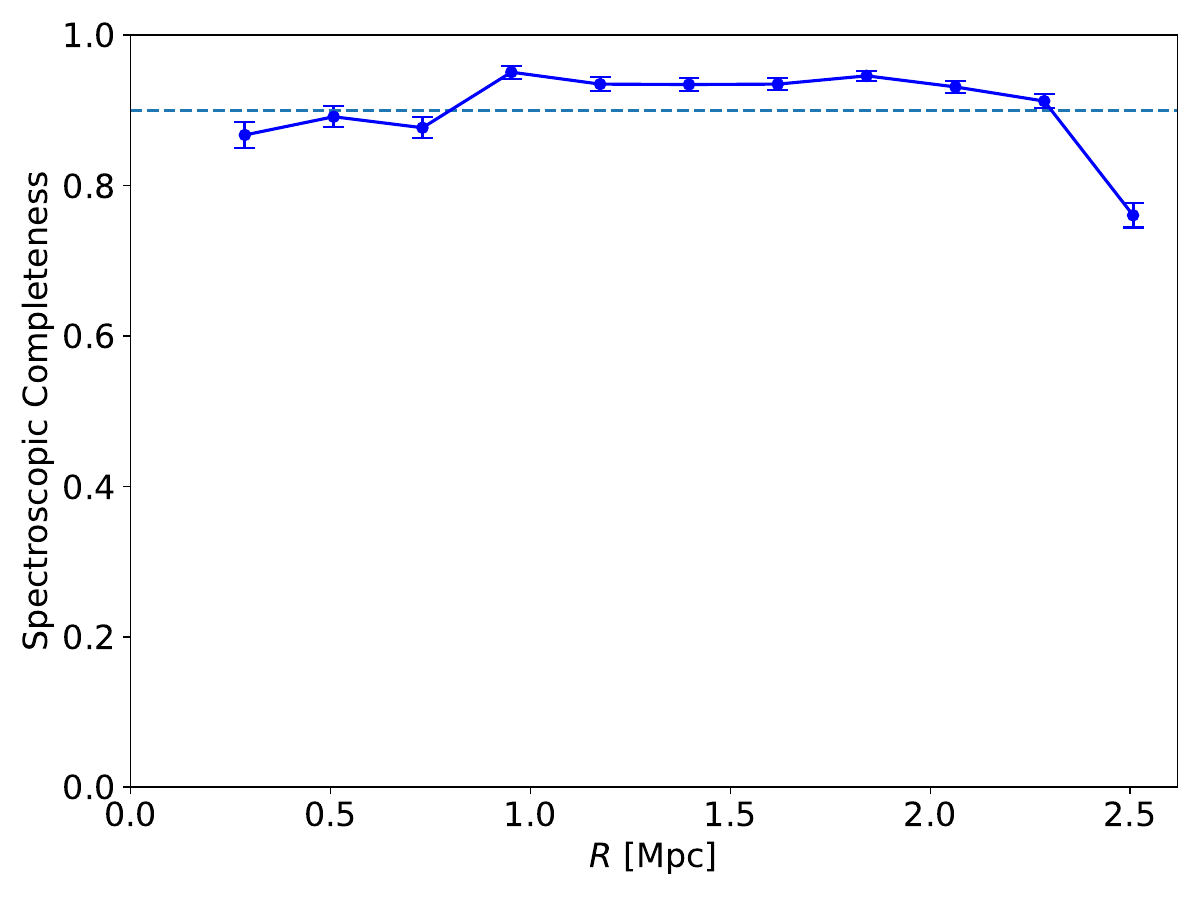}
    \caption{Spectroscopic completeness profile as a function of projected distance from the centre of the Coma cluster, with binomial error bars, up to $R=2.5 \, \mathrm{Mpc}$.}
    \label{fig:completeness}
\end{figure}

Figure \ref{fig:completeness} presents the spectroscopic completeness profile as a function of the projected distance from the cluster center. The spectroscopic completeness is evaluated as described in Sec. \ref{subsec:dataset} and shows a remarkable constancy up to $\approx$ 2.4 Mpc where it starts to decline. This radius corresponds to the edge of the DESI field of view (1.6 deg in radius) and marks the outer region where the target density is enhanced thanks to the highly dense DESI Science Verification program covering the Coma cluster.

\section{Impact of the assumed galaxy number density model}
\label{app:num_density_assump}

Our \textsc{MG-MAMPOSSt} runs all assumed a Hernquist model for the galaxy number density, with its scale radius $r_\nu$ (we drop the superscripts `H' or `NFW' in this appendix, since it is always clear what model we are using) kept as a free parameter with flat informative priors based on our preliminary fits of Sect.~\ref{num_density_methods}.

Here we aim to test, using \textsc{MG-MAMPOSSt}, whether:
\begin{enumerate}
\item[\textit{(i)}] the Hernquist model is indeed preferred over the NFW model for the galaxy number density profile according to the \textsc{MG-MAMPOSSt} likelihood;
\item[\textit{(ii)}] even when $r_\nu$ is treated as a free parameter, its best-fit value remains consistent with our initial estimate from Sect.~\ref{num_density_methods};
\item[\textit{(iii)}] the model with $r_\nu$ fixed at the best-fit value from Sect.~\ref{num_density_methods} is preferred relative to the case where $r_\nu$ is free.
\end{enumerate}

\noindent Table \ref{tab:model_comparison} shows the details of the runs performed, with relative priors and outcomes, to check \textit{(i)--(iii)}. The first two rows illustrate the comparison between NFW and Hernquist models: in these runs, $r_\nu$ is a free parameter, together with $r_{200\rm{c}}$, $r_{\rm s}$, $\mathcal{A}_0$ and $\mathcal{A_\infty}$. For $r_\nu$ we assumed informative flat priors coming from our best-fit estimate of Sect.~\ref{num_density_methods}, for $r_{200\rm{c}}$ and $r_{\rm s}$ we used broad and flat priors, as well as for the anisotropy parameters (same as Table \ref{tab:priors_flat}). We used the BIC and AIC criteria (Eq. \ref{eq:BIC} and Eq. \ref{eq:AIC}) to check the relative performance of the different models.
The results favour the Hernquist model with respect to the NFW model: we have tested this conclusion with several runs on different samples, including the three colour classes, and combinations of mass and anisotropy models, reaching the same outcome. This validates our first assumption and excludes any systematic errors caused by the employed galaxy number density model.

Also our second assumption is verified by the first two rows of Table \ref{tab:model_comparison}: in fact, the best-fit estimates of $r_\nu$ from \textsc{MG-MAMPOSSt} are very consistent with the fits performed in Sect.~\ref{num_density_methods}.  

The results of the third and fourth row of Table \ref{tab:model_comparison} show what happens if we treat $r_\nu$ as a fixed input parameter. In these runs, the flat priors for $r_\nu$ are very tight around the best-fit value from Sect. \ref{num_density_methods}; in practice, we mimic a fixed tracer scale radius by adopting a very narrow flat prior for $r_\nu$ and
reduce by hand the number of free parameters by unity. The reason why we do not consider $r_\nu$ as an actual fixed input parameter is that \textsc{MG-MAMPOSSt} uses a different expression for the fit depending on whether $r_\nu$ is a fixed or free parameter, invalidating the comparison of the likelihoods between the two cases. 
The results show that keeping $r_\nu$ 'fixed' with both NFW and Hernquist models produces an improvement of 7 BIC and 2 AIC units with respect to the free parameter case. This suggests that fixing $r_\nu$ to the value fitted outside \textsc{MG-MAMPOSSt}, instead of safely treating it as a free parameter, is a very valuable and statistically favoured strategy. This test will be very useful for future \textsc{MG-MAMPOSSt} runs.

\begin{table*}
\centering
\caption{MG-MAMPOSSt priors and results for changes on tracer number density models.}
\renewcommand{\arraystretch}{1.3} 
\begin{tabular}{lccccccc}
\hline
{Model} & {Min}\,$r_\nu \, (\mathrm{Mpc})$ & {Max}\,${r_\nu \, (\mathrm{Mpc})}$ & {Best fit} ${r_\nu \, (\mathrm{Mpc})}$ & ${k}$ & ${-\ln\mathcal{L}}$ & {BIC} & {AIC} \\
\hline
NFW  & 0.59 & 0.86 & $0.70^{+0.05}_{-0.06}$  & 5 & 12\,209.84  & 24\,455 & 24\,429 \\
Hernquist  & 1.41 & 2.05 & $1.70^{+0.13}_{-0.11}$  & 5 & 12\,207.59  & 24\,451 & 24\,425 \\
NFW   & 0.709 & 0.711  & $0.71$  & 4 & 12\,209.78  & 24\,448 & 24\,427 \\
Hernquist & 1.709 & 1.711 & 1.71 & 4 & 12\,207.51 & 24\,444 & 24\,423 \\
\hline
\end{tabular}
\tablefoot{ 
The columns report, in order, the model for the galaxy number density profile, the minimum and maximum limit of the flat employed priors, the best-fit values for $r_\nu$, the number of free parameters $k$, the minimum value for the negative log-likelihood, the BIC and AIC parameters. The number of free parameters for the third and fourth rows was decreased by one to mimic a constant value for $r_\nu$.}
\label{tab:model_comparison}
\end{table*}

\section{Surface number density profiles for different populations}
\label{app:sd_colors}

\begin{figure}[!h]
    \centering
    \includegraphics[width=0.9\linewidth]{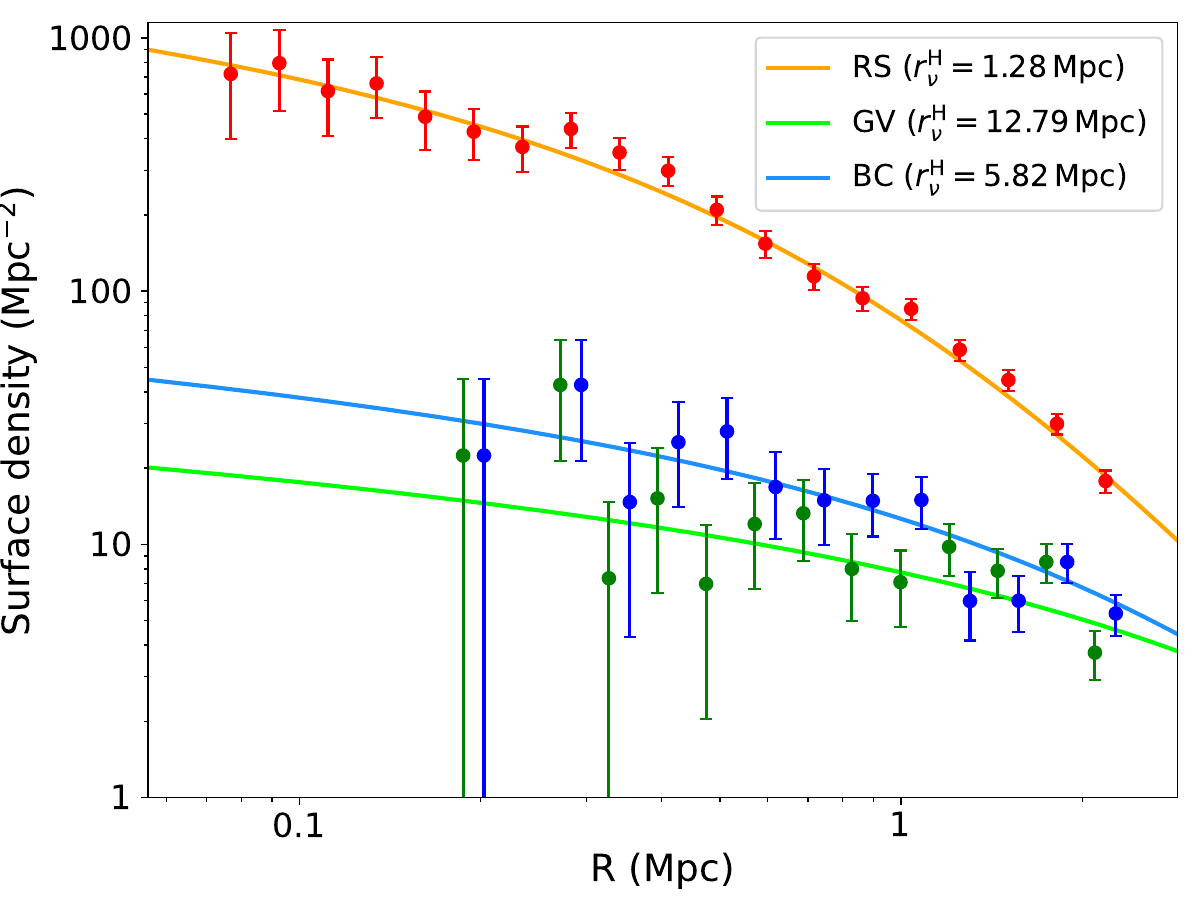}
    \caption{Surface density profiles of RS (red), GV (green), and BC (blue) galaxies in the Coma cluster fitted with Hernquist models with, respectively, $r_\nu^{\rm H} = 1.28\, \text{Mpc}, 12.79 \,\text{Mpc} \,\text{and}\, 5.82 \, \text{Mpc}$. The error bars correspond to the Poisson uncertainties.}
    \label{fig:surface_density_subsamples}
\end{figure}

Figure~\ref{fig:surface_density_subsamples} displays the surface density profiles of the different subsamples (RS, GV, BC) of galaxies, fitted with a Hernquist model. The low statistics of GV and BC samples are at the origin of the high uncertainties on the best-fit scale radii (Sect.~\ref{num_density_methods}).

\section{Multi-component mass profile} \label{app:mc}
\begin{figure}[!h]
    \centering
    \hspace{0.02\columnwidth}
    \includegraphics[width=0.8\columnwidth]{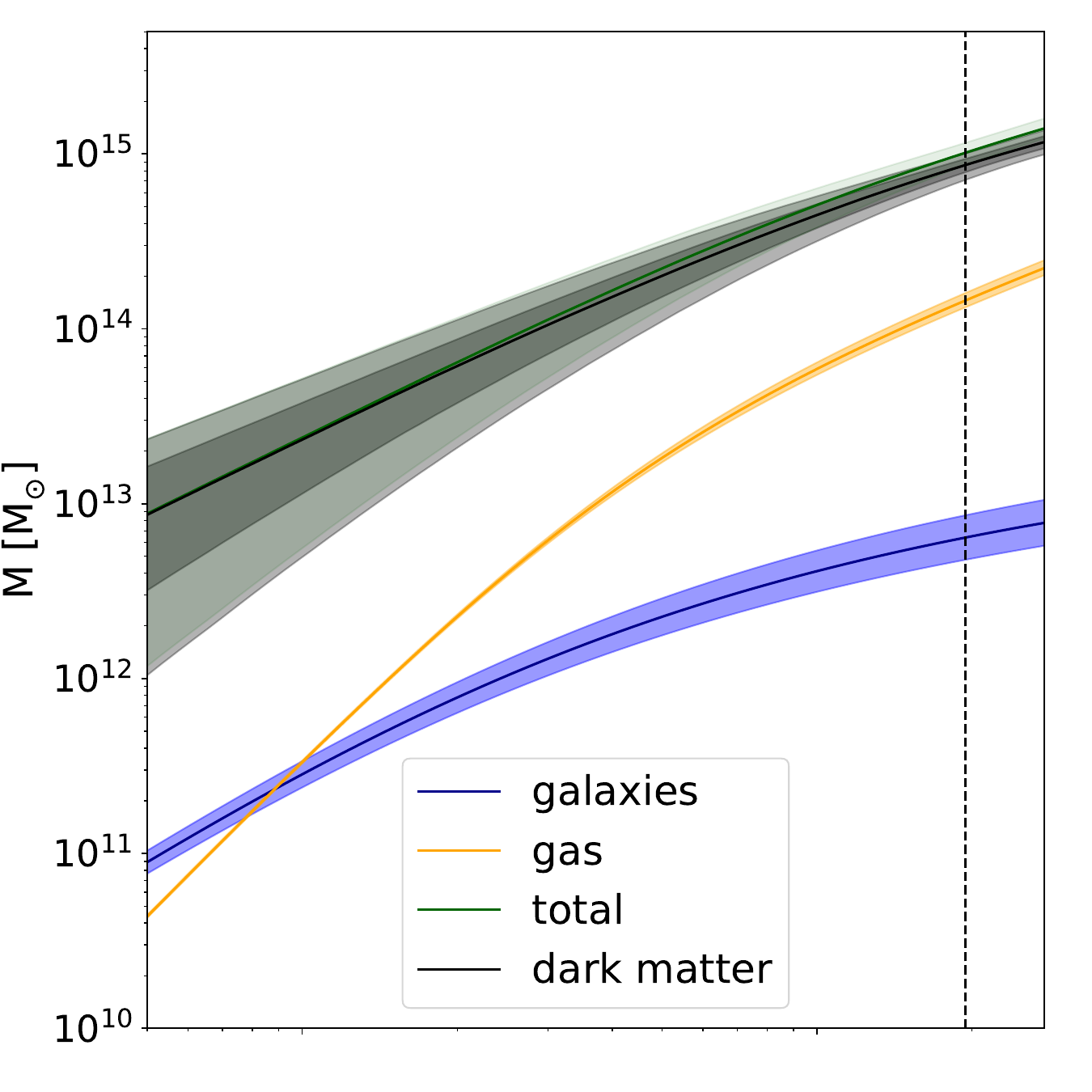}
    \includegraphics[width=0.79\columnwidth]{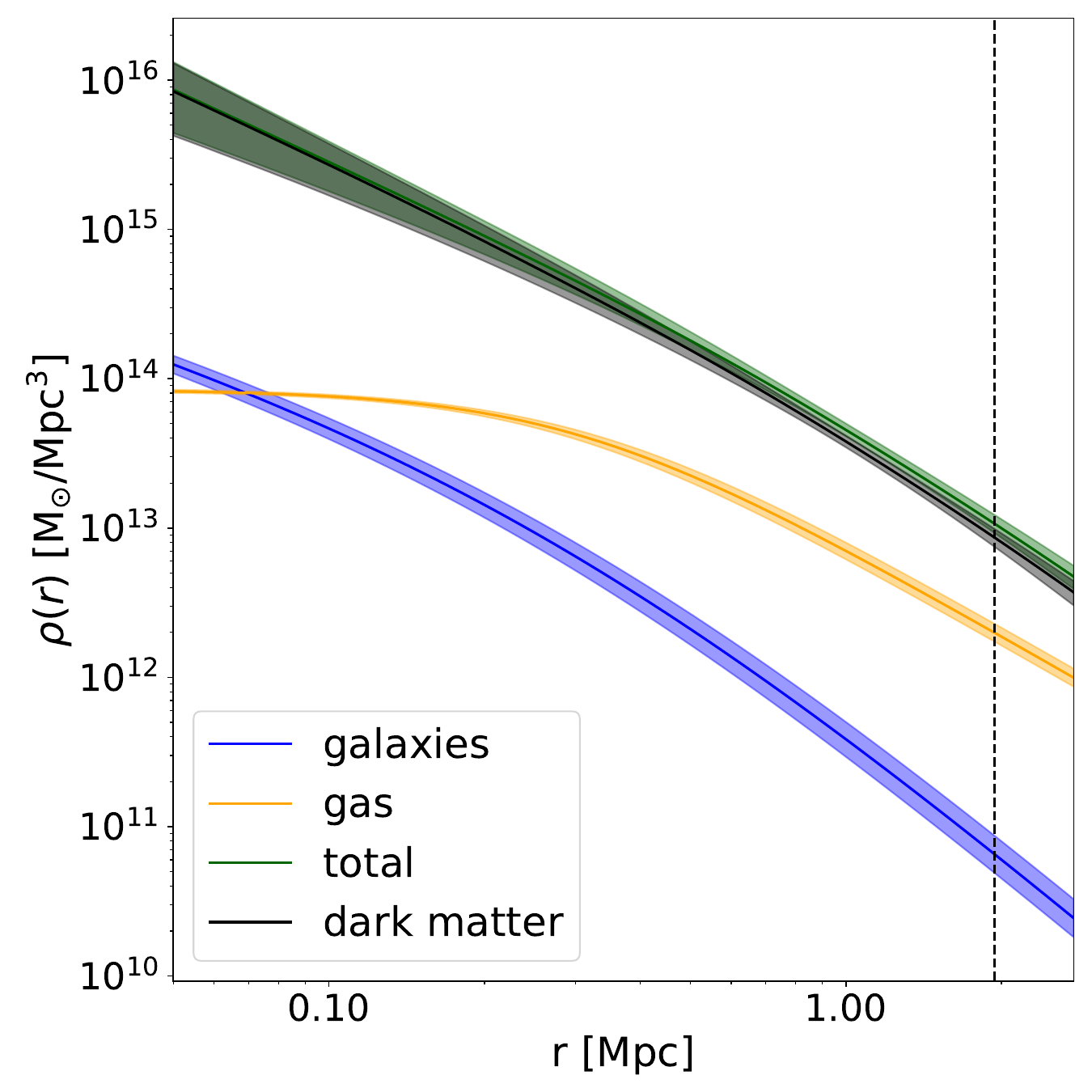}
    \caption{\textit{Top}: Total mass profile (green solid line) of Coma with its 95\% confidence interval (green shaded region), along with the mass of the different components: galaxies (blue),  gas (orange),  DM (black). The grey inner and outer shaded regions indicate the 68\% and 95\% confidence region of the DM profile, respectively. The black dashed vertical line corresponds to $r_{200\rm{c}}^{\rm DM} = 1.94^{+0.09}_{-0.06}$ Mpc. The orange and blue bands represent the 68\% limits of the gas and galaxies mass profiles. \textit{Bottom}: Corresponding density profiles, colour-coded as the left plot. Only the 68\% regions are shown for all the components.}
    \label{fig:multicompoent}
\end{figure}

In Fig.~\ref{fig:multicompoent} we plot the total mass profile of Coma obtained by using the sample of galaxies with no distinction in the three colour classes, in the projected radial range $R \in [0.10, 2.3]$ Mpc and assuming the multi-component mass model of Eq.~\eqref{eq:multic}. The grey bands and the black solid line indicate the radial profile of the diffuse DM component, parametrised as a gNFW. While the virial radius (and so the dark matter mass) is well constrained, $r_{200\rm{c}}^{\rm DM} = 1.94^{+0.09}_{-0.06}$ Mpc, the scale radius $r_{\rm s} = 0.51_{-0.21}^{+1.58}$ Mpc and the slope $\alpha=1.80^{+0.08}_{-0.48}$ are completely degenerate. This is not surprising; as shown in e.g.~\cite{Sartoris_2020} and \cite{Biviano_2023}, to break this degeneracy using kinematic analyses requires spectroscopic data of the stellar velocity dispersion in the BCG to be included in the fit.

\section{Mass and orbital anisotropy study for low- and high-mass galaxies}
\label{app:high-vs-low-mass}

As mentioned in Sect. \ref{subsec:diff_sample_mamposst}, we split the full sample in two equal population (662 members) stellar mass bins, with the threshold at $\log (M_*/M_\odot) = 9.1$, to check differences in the estimated mass and anisotropy parameters. RS galaxies are the most evenly distributed between the two bins, with 46\% below the mass threshold and 54\% above it. GV galaxies show a similarly balanced distribution, with 55\% with low mass and 45\% with high mass. In contrast, BC galaxies are strongly skewed towards the low-mass regime, where 75\% of them are found. The colour composition of the two stellar mass bins can be found in Table \ref{tab:composition_colors}. 

\begin{table}
\centering
\renewcommand{\arraystretch}{1.3} 
\caption{Fraction of RS, GV and BC galaxies in the two stellar-mass bins.}
\begin{tabular}{lcc}
\hline
Population & $\log (M_\star/M_\odot) < 9.1$ & $\log (M_\star/M_\odot) > 9.1$ \\
\hline
RS & 71.6\% & 85.3\% \\
GV & 10.9\% & 8.9\% \\
BC & 17.5\% & 5.8\% \\
\hline
\label{tab:composition_colors}
\end{tabular}
\end{table}

We first ran \textsc{MG-MAMPOSSt} with flat priors (same as Table \ref{tab:priors_flat}) to independently estimate the mass parameters and then we employed Gaussian priors on the mass from the joint analysis of Sect. \ref{full_sample_mamposst} to compare the anisotropy profiles. The best-fit parameters are reported in Tables \ref{tab:massbins_results} and \ref{tab:massbins_results_withpriors}. Note that we have repeated the galaxy number density profile fit for these new subclasses (as in Sect. \ref{num_density_methods}), to ensure that Gaussian priors on $r_\nu^{\rm H}$ are appropriate, finding the best-fit values of $r_\nu^{\rm H}=2.20_{-0.19}^{+0.24} \, \mathrm{Mpc}$ for the low-mass and $r_\nu^{\rm H}=1.52_{-0.14}^{+0.18} \, \mathrm{Mpc}$ for the high-mass galaxies.

\begin{table}
\centering
\caption{Best fit (maximum posterior) parameters for low-mass and high-mass galaxies.}
\renewcommand{\arraystretch}{1.3}
\begin{tabular}{lccc}
\hline
{Parameter} & {Low-mass} & {High-mass}  \\
\hline
       
       $r_{200\rm{c}} \, (\mathrm{Mpc})$ & $2.23_{-0.12}^{+0.10}$ & $2.13_{-0.09}^{+0.06}$  \\
       $r_{\rm s} \, (\mathrm{Mpc})$ & $0.82_{-0.52}^{+0.55}$ & $0.58_{-0.35}^{+0.10}$ \\
       $r_\nu^{\rm H} \, (\mathrm{Mpc})$ & $2.09_{-0.28}^{+0.22}$ & $1.35_{-0.10}^{+0.20}$  \\
       $\mathcal{A}_0$ & 
       $1.42_{-0.51}^{+0.63}$ & $1.14_{-0.42}^{+0.92}$  \\
       $\mathcal{A}_\infty$ & $1.20_{-0.32}^{+0.25}$ & $1.42_{-0.47}^{+0.14}$  \\
          
\hline
\end{tabular}
\tablefoot{The employed mass and anisotropy models are NFW and gOM, respectively. The assumed priors, for these runs, are broad and flat, except for $r_\nu^{\rm H}$, for which we considered informative priors based on the best-fit values for each subsample obtained with the methodology of Sect. \ref{num_density_methods}.}
\label{tab:massbins_results}
\end{table}

\begin{figure*}
    \centering
    \includegraphics[width=0.3565\linewidth]{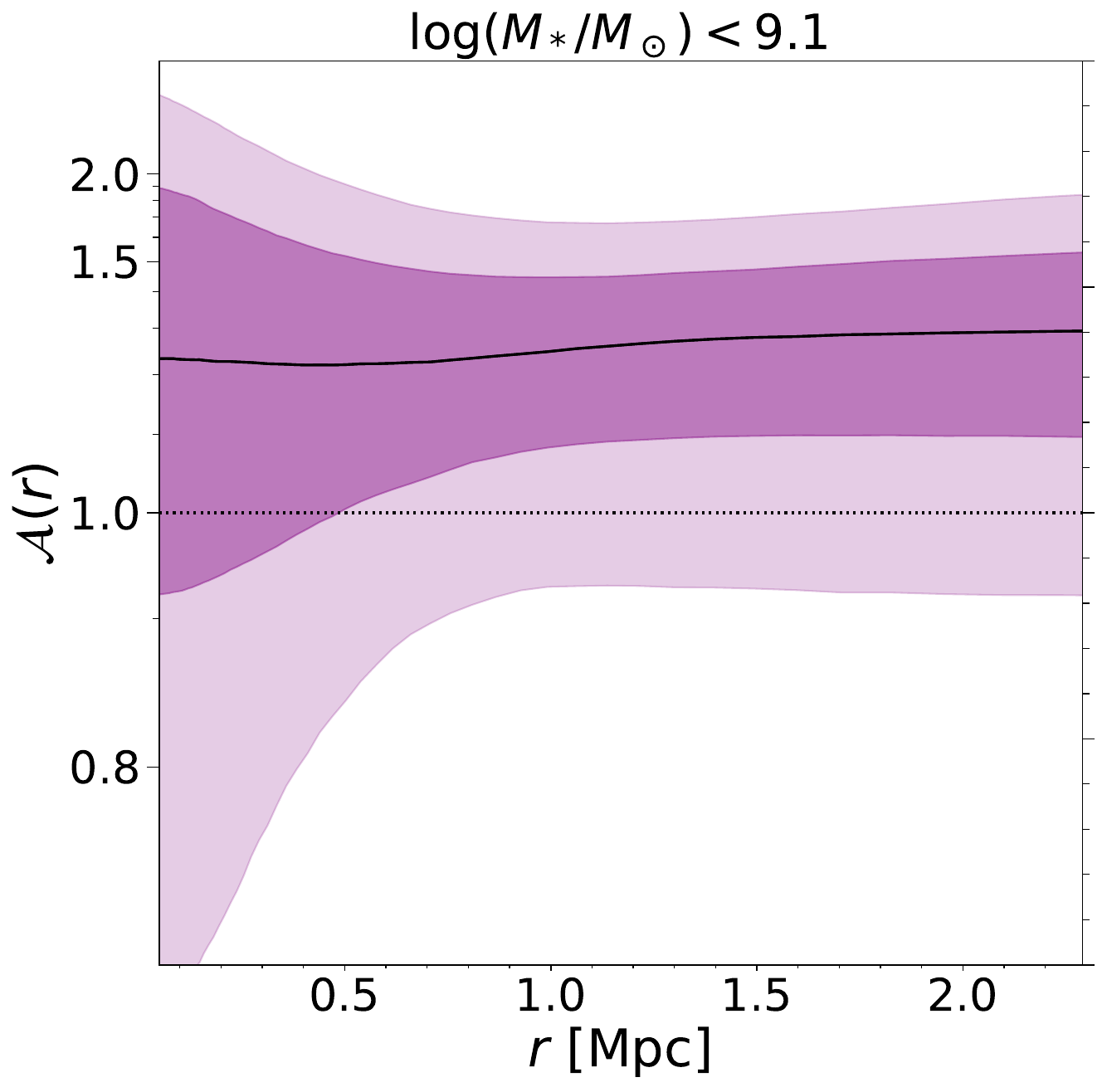}
    \includegraphics[width=0.3615\linewidth]{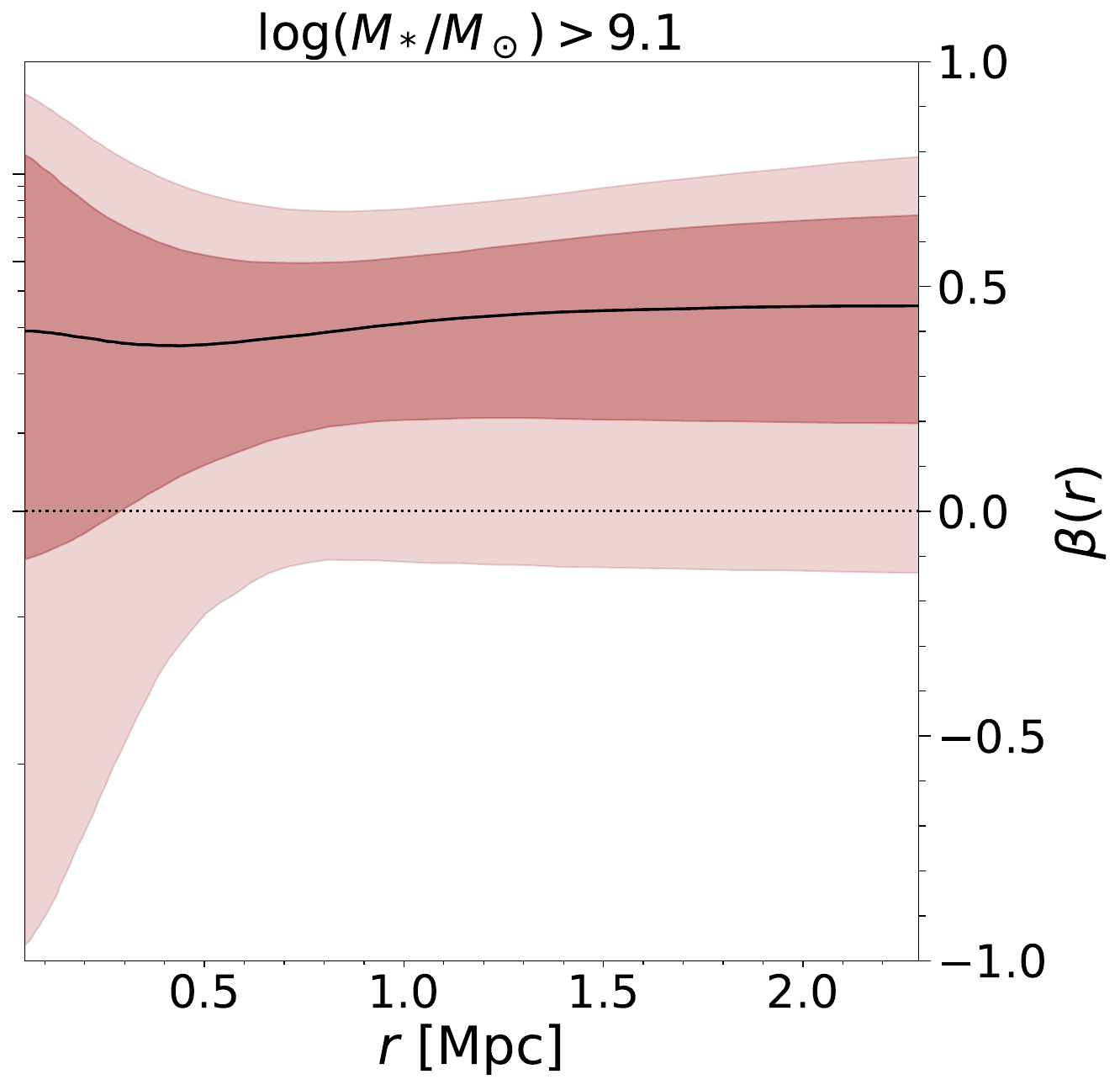}

    \caption{Estimated anisotropy profiles in NFW gOM runs with Gaussian priors on the mass (coming from the joint results of Sect. \ref{full_sample_mamposst}) for low-mass (\textit{left}) and high-mass (\textit{right}) galaxies, separately. The coloured and shaded regions indicate 1$\,\sigma$ and 2$\,\sigma$ confidence intervals, respectively. Solid curves are the median profiles of each MCMC chain, while horizontal dotted lines highlight isotropic orbits.}
    \label{fig:anis_massbins}
\end{figure*}

As expected, the mass parameters ($r_{\rm s}$ and $r_{200\rm{c}}$) estimated from the low-mass sample are slightly higher than those obtained from the high-mass one (still consistent within the uncertainties), given the contribution of the vast majority of BC galaxies. Both anisotropy profiles (Fig.~\ref{fig:anis_massbins}) show a predominant radial component, with no significant differences between the two populations.

\begin{table}
\centering
\caption{Same as Table \ref{tab:massbins_results} but for the runs with Gaussian priors on $r_{200\rm{c}}$ and $r_{\rm{s}}$ from the joint analysis of Sect. \ref{full_sample_mamposst}.}
\renewcommand{\arraystretch}{1.3}
\begin{tabular}{lccc}
\hline
{Parameter} & {Low-mass} & {High-mass} \\
\hline
       
       $r_{200\rm{c}} \, (\mathrm{Mpc})$ & $2.11_{-0.03}^{+0.06}$ & $2.08_{-0.03}^{+0.05}$  \\
       $r_{\rm s} \, (\mathrm{Mpc})$ & $0.84_{-0.35}^{+0.14}$ & $0.56_{-0.22}^{+0.20}$  \\
       $r_\nu^{\rm H} \, (\mathrm{Mpc})$ & $2.16_{-0.12}^{+0.18}$ & $1.46_{-0.09}^{+0.11}$  \\
       $\mathcal{A}_0$ & 
       $1.05_{-0.28}^{+0.51}$ & $1.07_{-0.32}^{+0.61}$  \\
       $\mathcal{A}_\infty$ & $1.46_{-0.44}^{+0.06}$ & $1.36_{-0.36}^{+0.27}$  \\
          
\hline
\end{tabular}
\tablefoot{Gaussian priors on $r_\nu^{\rm H}$ are based on the estimates obtained outside \textsc{MG-MAMPOSSt}, while those for $\mathcal{A}_0$ and $\mathcal{A}_\infty$ are flat (same as Table \ref{tab:priors_flat}).}
\label{tab:massbins_results_withpriors}
\end{table}

\end{document}